\documentclass[10pt,letterpaper]{article}
\usepackage[top=0.85in,left=2.75in,footskip=0.75in]{geometry}

\usepackage{amsmath,amssymb}

\usepackage{changepage}

\usepackage{textcomp,marvosym}

\usepackage{cite}

\usepackage{nameref,hyperref}

\usepackage[right]{lineno}

\usepackage[nopatch=eqnum]{microtype}
\DisableLigatures[f]{encoding = *, family = * }

\usepackage[table]{xcolor}
\hypersetup{
    colorlinks,
    linkcolor={red!50!black},
    citecolor={blue!50!black},
    urlcolor={blue!80!black}
}

\usepackage{array}

\newcolumntype{+}{!{\vrule width 2pt}}

\newlength\savedwidth

\newcommand\thickhline{\noalign{\global\savedwidth\arrayrulewidth\global\arrayrulewidth 2pt}%
\hline
\noalign{\global\arrayrulewidth\savedwidth}}

\raggedright
\usepackage[aboveskip=1pt,labelfont=bf,labelsep=period,justification=raggedright,singlelinecheck=off]{caption}

\makeatletter
\renewcommand{\@biblabel}[1]{\quad#1.}
\makeatother

\usepackage{listings}

\usepackage{lastpage,fancyhdr,graphicx}
\usepackage{epstopdf}
\fancyheadoffset[L]{2.25in}
\fancyfootoffset[L]{2.25in}
\usepackage[inline]{enumitem}

\newlist{inlineenum}{enumerate*}{1}
\setlist[inlineenum]{label=\roman*)}

\newcommand{\termsorg}[1]{\href{https://schema.org/#1}{\color{black}{\emph{s:#1}}}}
\newcommand{\termbioschemas}[1]{\href{https://bioschemas.org/#1}{\color{black}{\emph{bioschemas:#1}}}}
\newcommand{\termbsp}[1]{\href{https://bioschemas.org/properties/#1}{\color{black}{\emph{bsp:#1}}}}
\newcommand{\termwfrun}[1]{\href{https://w3id.org/ro/terms/workflow-run\##1}{\color{black}{\emph{wfrun:#1}}}}

\begin{document}
\vspace*{0.2in}

\begin{flushleft}
{\Large
\textbf\newline{Recording provenance of workflow runs with RO-Crate} 
}
\newline
\\

Simone Leo\textsuperscript{1*},
Michael R. Crusoe\textsuperscript{2,3,4},
Laura Rodríguez-Navas\textsuperscript{5}, 
Raül Sirvent\textsuperscript{5}, 
Alexander Kanitz\textsuperscript{6,7}, 
Paul De Geest\textsuperscript{8}, 
Rudolf Wittner\textsuperscript{9,10,11}, 
Luca Pireddu\textsuperscript{1}, 
Daniel Garijo\textsuperscript{12}, 
José M. Fernández\textsuperscript{5}, 
Iacopo Colonnelli\textsuperscript{13}, 
Matej Gallo\textsuperscript{9}, 
Tazro Ohta\textsuperscript{14,15}, 
Hirotaka Suetake\textsuperscript{16}, 
Salvador Capella-Gutierrez\textsuperscript{5}, 
Renske de Wit\textsuperscript{2}, 
Bruno P. Kinoshita\textsuperscript{5}, 
Stian Soiland-Reyes\textsuperscript{17,18}
\\
\bigskip
\textbf{1} Center for Advanced Studies, Research, and Development in Sardinia (CRS4), Pula (CA), Italy
\\
\textbf{2} Vrije Universiteit Amsterdam, Amsterdam, The Netherlands
\\
\textbf{3} DTL Projects, The Netherlands
\\
\textbf{4} Forschungszentrum Jülich, Germany
\\
\textbf{5} Barcelona Supercomputing Center, Barcelona, Spain
\\
\textbf{6} Biozentrum, University of Basel, Basel, Switzerland
\\
\textbf{7} Swiss Institute of Bioinformatics, Lausanne, Switzerland
\\
\textbf{8} VIB Data Core, Gent, Belgium
\\
\textbf{9} Faculty of Informatics, Masaryk University, Brno, Czech Republic
\\
\textbf{10} Institute of Computer Science, Masaryk University, Brno, Czech Republic
\\
\textbf{11} BBMRI-ERIC, Graz, Austria
\\
\textbf{12} Ontology Engineering Group, Universidad Politécnica de Madrid, Madrid, Spain
\\
\textbf{13} Computer Science Department, Università degli Studi di Torino, Torino, Italy
\\
\textbf{14} Database Center for Life Science, Joint Support-Center for Data Science Research, Research Organization of Information and Systems, Shizuoka, Japan
\\
\textbf{15} Institute for Advanced Academic Research, Chiba University, Chiba, Japan
\\
\textbf{16} Sator, Incorporated, Tokyo, Japan
\\
\textbf{17} Department of Computer Science, The University of Manchester, Manchester, United Kingdom
\\
\textbf{18} Informatics Institute, University of Amsterdam, Amsterdam, The Netherlands
\\
\bigskip

%
%





* simone.leo@crs4.it (SL)

\end{flushleft}
\section*{Abstract}
Recording the provenance of scientific computation results is key to the support of traceability, reproducibility and quality assessment of data products.
Several data models have been explored to address this need, providing representations of workflow plans and their executions as well as means of packaging the resulting information for archiving and sharing.
However, existing approaches tend to lack interoperable adoption across workflow management systems.
In this work we present Workflow Run RO-Crate, an extension of RO-Crate (Research Object Crate) and Schema.org to capture the provenance of the execution of computational workflows at different levels of granularity and bundle together all their associated objects (inputs, outputs, code, etc.).
The model is supported by a diverse, open community that runs regular meetings, discussing development, maintenance and adoption aspects.
Workflow Run RO-Crate is already implemented by several workflow management systems, allowing interoperable comparisons between workflow runs from heterogeneous systems.
We describe the model, its alignment to standards such as W3C PROV, and its implementation in six workflow systems.
Finally, we illustrate the application of Workflow Run RO-Crate in two use cases of machine learning in the digital image analysis domain.

\linenumbers


\section{Introduction}\label{introduction}

A crucial part of scientific research is recording the provenance of its outputs.
The W3C PROV standard defines provenance as ``a record that describes the people, institutions, entities, and activities involved in producing, influencing, or delivering a piece of data or a thing''~\cite{Moreau 2013}.
Provenance is instrumental to activities such as traceability, reproducibility,
accountability, and quality assessment~\cite{Herschel 2017}.
The constantly growing size and complexity of scientific datasets and the analysis that is required to extract useful information from them has made science increasingly dependent on advanced automated processing techniques in order to get from experimental data to final results~\cite{Himanen 2019, Gauthier 2019, Huntingford 2019}.
Consequently, a large part of the provenance information for scientific outputs consists of descriptions of complex computer-aided data processing steps. This data processing is often expressed as workflows -- i.e., high-level applications that coordinate multiple tools and manage intermediate outputs in order to produce the final results.

In order to homogenise the collection and interchange of provenance records, the W3C consortium proposed a standard for representing provenance in the Web (PROV ~\cite{Moreau 2013}), along with the PROV ontology (PROV-O)~\cite{Lebo 2013}, an OWL~\cite{W3C OWL Working Group 2012} representation of PROV. 
PROV-O has been widely extended for workflows (e.g., D-PROV~\cite{Missier 2013}, ProvONE~\cite{Cuevas-Vicenttin 2016}, OPMW~\cite{Garijo 2011} (Open Provenance Model for Workflows), P-PLAN~\cite{Garijo 2012}), where provenance information is collected in two main forms: prospective and retrospective~\cite{Freire 2008}. \emph{Prospective provenance} -- the execution plan -- is essentially the workflow itself: it includes a machine-readable specification with the processing steps to be performed and the data and software dependencies to carry out each computation.
\emph{Retrospective provenance} refers to what actually happened during an execution -- i.e.~what were the values of the input parameters, which outputs were produced, which tools were executed, how much time did the execution take, whether the execution was successful or not, etc.
Retrospective provenance may be represented at different levels of abstraction, depending on the information that is available and/or required: a workflow execution may be interpreted
\begin{inlineenum}
\item as a single end-to-end activity,
\item as a set of individual execution of workflow steps, or
\item by going a step further and indicating how each step is divided into sub-processes when a workflow is deployed in a cluster.
\end{inlineenum}
Various workflow management systems, such as WINGS~\cite{Gil 2011} (Workflow INstance Generation and Specialization) and VisTrails~\cite{Scheidegger 2008,Costa 2013}, have adopted PROV and its PROV-O representation  to lift the burden of provenance collection from tool users and developers~\cite{Atkinson 2017,Perez 2018}.

D-PROV, PROV-ONE, OPMW, P-PLAN propose representations of workflow plans and their respective executions, taking into account the features of the workflow systems implementing them (e.g., hierarchical representations, sub-processes, etc.).
Other data models, such as \emph{wfprov} and \emph{wfdesc}~\cite{Belhajjame 2015}, go a step further by considering not only the link between plans and executions, but also how to package the various artefacts as a Research Object (RO)~\cite{Bechhofer 2013} to improve metadata interoperability and document the context of a digital experiment.

However, while these models address some workflow provenance representation issues, they have two main limitations: first, the extensions of PROV are not directly interoperable because of differences in their granularities or different assumptions in their workflow representations; second, their support from Workflow Management Systems (WMS) is typically one system per model.  An early approach to unify and integrate workflow provenance traces across WMSs was the Workflow Ecosystems through STandards (WEST)~\cite{Garijo 2014}, which used WINGS to build workflow templates and different converters. In all of these workflow provenance models, the emphasis is on the workflow execution structure as a directed graph, with only partial references for the data items.
The REPRODUCE-ME ontology~\cite{Samuel 2022} extended PROV and P-PLAN to explain the overall scientific process with the experimental context including real life objects (e.g. instruments, specimens) and human activities (e.g. lab protocols, screening), demonstrating provenance of individual Jupyter Notebook cells~\cite{Samuel 2018} and highlighting the need for provenance also where there is no workflow management system.

More recently, interoperability has been partially addressed by Common Workflow Language Prov (CWLProv)~\cite{Khan 2019}, which represents workflow enactments as research objects serialised according to the Big Data Bag approach~\cite{Chard 2016}.
The resulting format is a folder containing several data and metadata files~\cite{Soiland-Reyes 2018}, expanding on the Research Object Bundle approach of Taverna~\cite{Soiland-Reyes 2016}.
CWLProv also extends PROV with a representation of executed processes (activities), their inputs and outputs (entities) and their executors (agents), together with their Common Workflow Language (CWL) specification~\cite{Crusoe 2022} -- a standard workflow specification adopted by at least a dozen different workflow systems~\cite{cwl-implementations}. Although CWLProv includes prospective provenance as a \emph{plan}
within PROV (based on the \emph{wfdesc} model), in practice its implementation does not include tool definitions or file formats.
Thus, for CWLProv consumers to reconstruct the full prospective provenance for understanding the workflow, they would also need to inspect the separate workflow definition in the native language of the workflow management system.
Additionally, the CWLProv RO may include several other metadata files and PROV serialisations conforming to different formats, complicating its generation and consumption.

As for granularity, CWLProv proposes multiple levels of provenance~\cite[Figure 2]{Khan 2019}, from Level 0 (capturing workflow definition) to Level 3 (domain-specific annotations).
In practice, the CWL reference implementation \emph{cwltool}~\cite{Amstutz 2023} and the corresponding CWLProv specification~\cite{Soiland-Reyes 2018} record provenance details of all task executions together with the intermediate data and any nested workflows (CWLProv level 2). This level of granularity requires substantial support from the workflow management system implementing the CWL specification, resulting appropriate for workflow languages where the execution plan, including its distribution among the various tasks, is well known in advance.
However, it can be at odds with other systems where the execution is more dynamic, depending on the verification of specific runtime conditions, such as the size and distribution of the data (e.g., COMPSs~\cite{Lordan 2014}).
This design makes the implementation of CWLProv challenging, which the authors suspect may be one of the main causes for the low adoption of CWLProv (at the time of writing the format is supported only by cwltool).
Finally, being based on the PROV model, CWLProv is highly focused on the interaction between agents, processes and related entities, while support for contextual metadata (such as workflow authors, licence or creation date) in the Research Object Bundle is limited~\cite{rob-context} and stored in a separate manifest file, which includes the data identifier mapping to filenames.
A project that uses serialised Research Objects similar to those used by CWLProv is Whole Tale~\cite{Chard 2019}, a web platform with a focus on the narrative around scientific studies and their reproducibility, where the serialised ROs are used to export data and metadata from the platform. In contrast, our work is primarily focused on the ability to capture the provenance of computational workflow execution including its data and executable workflow definitions.

RO-Crate~\cite{Soiland-Reyes 2022a} is an approach for packaging research data together with their metadata and associated resources. RO-Crate extends Schema.org~\cite{Guha 2015}, a popular vocabulary for describing resources on the Web.
In its simplest form, an RO-Crate is a directory structure that contains a single JSON-LD~\cite{w3-json-ld} metadata file at the top level.
The metadata file describes all entities stored in the RO-Crate along with their relationships, and it is both machine-readable and human-readable.
RO-Crate is general enough to be able to describe any dataset, but can also be made as specific as needed through the use of extensions called \emph{profiles}. Profiles describe ``a set of conventions, types and properties that one minimally can require and expect to be present in that subset of RO-Crates"~\cite{profiles-ro-crate}. 
The broad set of types and properties from Schema.org, complemented by a few additional terms from other vocabularies, make the RO-Crate model a candidate for expressing a wide range of contextual information that complements and enriches the core information specified by the profile.
This information may include, among others, the workflow authors and their affiliations, associated publications, licensing information, related software, etc.
This approach is used by WorkflowHub~\cite{Goble 2021}, a workflow-system-agnostic workflow registry which specifies a Workflow RO-Crate profile~\cite{Bacall 2022} to gather the workflow definition with such metadata in an archived RO-Crate.

In this work, we present \textbf{Workflow Run RO-Crate} (WRROC), an extension of RO-Crate for representing computational workflow execution provenance.
Our main contributions include:
\begin{itemize}
\item   a collection of RO-Crate profiles to represent and package both the prospective and the retrospective provenance of a computational workflow run in a way that is machine-actionable~\cite{Batista 2022}, independently of the specific workflow language or execution system, and including support for re-execution;
\item   implementations of this new model in six workflow management systems and in one conversion tool;
\item   a mapping of our profiles against the W3C PROV-O Standard using the Simple Knowledge Organisation System (SKOS)~\cite{Isaac 2009}.
\end{itemize}

To foster usability, the profiles are characterised by different levels of detail, and the set of mandatory metadata items is kept to a minimum in order to ease the implementation.
This flexible approach increases the model's adaptability to the diverse landscape of WMSs used in practice.
The base profile, in particular, is applicable to any kind of computational process, not necessarily described in a formal workflow language.
All profiles are supported and sustained by the Workflow Run RO-Crate community, which meets regularly to discuss extensions, issues and new implementations.

The rest of this work is organised as follows: we first describe the Workflow Run RO-Crate profiles in Section~\ref{the-workflow-run-ro-crate-profiles}; we then illustrate implementations in Section~\ref{implementations} and usage examples in Section~\ref{exemplary-use-cases}; finally, we include a discussion in Section~\ref{discussion} and we conclude the paper with our plans for future work in Section~\ref{conclusion}.

\section{The Workflow Run RO-Crate profiles}\label{the-workflow-run-ro-crate-profiles}

RO-Crate profiles are extensions of the base RO-Crate specification that describe how to represent the classes and relationships that appear in a specific domain or use case.
An RO-Crate conforming to a profile is not just machine-readable, but also machine-actionable, as a digital object whose type is represented by the profile itself~\cite{Soiland-Reyes 2022b}.

The Workflow Run RO-Crate profiles are the main outcome of the activities of the Workflow Run RO-Crate Community~\cite{wrroc-site}, an open working group that includes workflow users and developers, WMS users and developers, and researchers and software engineers interested in workflow execution provenance and Findable, Accessible, Interoperable and Reusable (FAIR) approaches for data and software.
%
One of the first steps in the development of the Workflow-Run RO-Crate profiles was to compile a list of requirements to be addressed by the model from all interested participants, in the form of \textit{competency questions}~(CQs)~\cite{wrroc-cqs}.
The process also included reviewing existing state of the art models, such as wfprov~\cite{Belhajjame 2015}, ProvONE~\cite{Cuevas-Vicenttin 2016} or OPMW~\cite{Garijo 2011}. The result was the definition of 11 CQs capturing requirements which span a broad application scope and consider different levels of provenance granularity.
Each requirement was supported by a rationale and linked to a GitHub issue to drive the public discussion forward. When a requirement was addressed, related changes were integrated into the profiles and the relevant issue was closed. All the original issues are now closed, and the profiles have had five official releases on Zenodo~\cite{WRROC 2024a, WRROC 2024b, WRROC 2024c}.
%
The target of several of the original CQs evolved during profile development, as the continuous discussion within the community highlighted the main points to be addressed. This continuous process is reflected in the corresponding issues and pull requests in the community's GitHub repository. The final implementation of the CQs in the profiles is validated with SPARQL queries that can be run on RO-Crate metadata samples, also available on the GitHub repository~\cite{cqs-sparql-queries}.

As requirements were being defined, it became apparent that one single profile would not have been sufficient to cater for all possible usage scenarios.
In particular, while some use cases required a detailed description of all computations orchestrated by the workflow, others were only concerned with a ``black box'' representation of the workflow and its execution as a whole (i.e., whether the workflow execution as a whole was successful and which results were obtained).
Additionally, some computations involve a data flow across multiple applications that are executed without the aid of a WMS and thus are not formally described in a standard workflow language.
These observations led to the development of three profiles:
\begin{enumerate}
    \item \textit{Process Run Crate},
 to describe the execution of one or more tools that contribute to a computation;
    \item \textit{Workflow Run Crate},
 to describe a computation orchestrated by a predefined workflow; 
    \item \textit{Provenance Run Crate},
 to describe a workflow computation including the internal details of individual step executions.
\end{enumerate} 

In the rest of this section we describe each of these profiles in detail. We use the term ``class'' to refer to a type as defined in RDF(s) and ``entity'' to refer to an instance of a class. We use italics to denote the properties and classes in each profile: these are defined in the RO-Crate JSON-LD context~\cite{roc-context}, which extends Schema.org with terms from the Bioschemas~\cite{Gray 2017} ComputationalWorkflow profile~\cite{computational-workflow-profile} and other vocabularies.
Note that terms coming from Bioschemas are not specific to the life sciences.
We also developed a dedicated term set~\cite{wrroc-terms} to represent concepts that are not captured by terms in the RO-Crate context. New terms are defined in RDF(s) following Schema.org guidelines (i.e., using \emph{domainIncludes} and \emph{rangeIncludes} to define domains and ranges of properties). 
In the rest of the text and images, the following prefixes are used to represent the corresponding namespaces:
\begin{tabular}{rcl}
\emph{s:}         & $\rightarrow$ & \url{https://schema.org/} \\
\emph{bioschemas:}& $\rightarrow$ & \url{https://bioschemas.org/} \\
\emph{bsp:}       & $\rightarrow$ & \url{https://bioschemas.org/properties/} \\
\emph{wfrun:}     & $\rightarrow$ & \url{https://w3id.org/ro/terms/workflow-run\#} \\
\end{tabular}

\subsection{Process Run Crate}\label{process-run-crate}

The Process Run Crate profile~\cite{WRROC 2024a} contains specifications to describe the execution of one or more software applications that contribute to the same overall computation, but are not necessarily coordinated by a top-level workflow or script (e.g. when executed manually by a human, one after the other as intermediate datasets become available).

\begin{figure}[!h]
\includegraphics[width=26em]{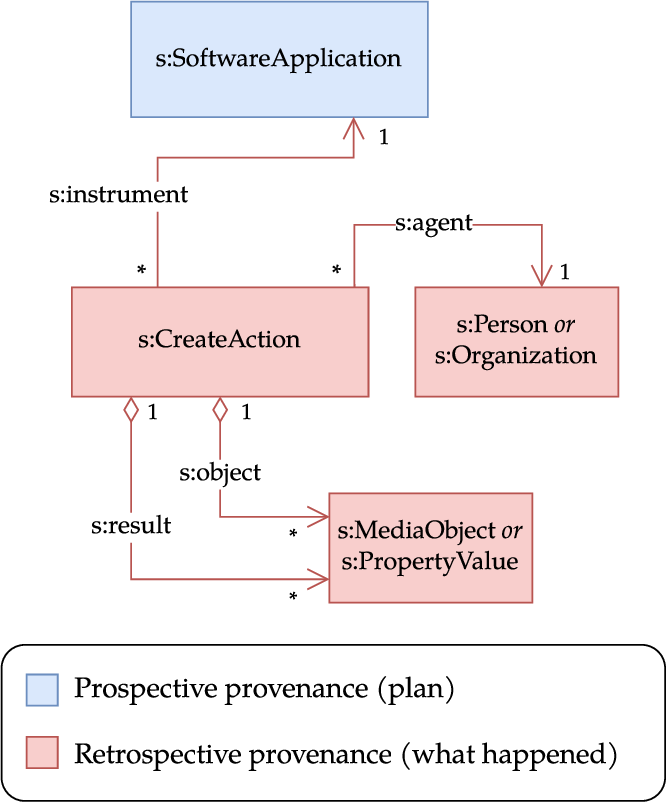}
\caption{{\bf UML class diagram for Process Run Crate.}
The central class is the \termsorg{CreateAction}, which represents the execution of an application.
It links to the application itself via \termsorg{instrument}, to the entity that executed it via \termsorg{agent}, and to its inputs and outputs via \termsorg{object}
and \termsorg{result}, respectively.
In this and following figures, classes and properties are shown with prefixes to indicate their origin. 
Some inputs (and, less commonly, outputs) are not stored as files or directories, but passed to the application (e.g., via a command line interface) as values of various types (e.g., a number or string). In this case, the profile recommends a representation via \termsorg{PropertyValue}.
For simplicity, we left out the rest of the RO-Crate structure (e.g. the root \termsorg{Dataset}), and attributes (e.g. \termsorg{startTime}, \termsorg{endTime}, \termsorg{description}, \termsorg{actionStatus}).
In this UML class notation, diamond $\Diamond$ arrows indicate aggregation and regular arrows indicate references, $*$ indicates zero or more occurrences, $1$ means single occurrence.  
}
\label{fig:process_crate_er}
\end{figure}

The Process Run Crate is the basis for all profiles in the WRROC collection. It specifies how to describe the fundamental classes involved in a computational run: \begin{inlineenum}
\item a software application represented by a \termsorg{SoftwareApplication}, \termsorg{SoftwareSourceCode} or \termbioschemas{ComputationalWorkflow} class; and
\item its execution, represented by a \termsorg{CreateAction} class, and linking to the application via the \termsorg{instrument} property.
\end{inlineenum}
Other important properties of the
\termsorg{CreateAction} class are \termsorg{object}, which links to the action's inputs, and \termsorg{result}, which links to its outputs.
The time the execution started and ended can be provided, respectively, via the
\termsorg{startTime} and \termsorg{endTime} properties.
The \termsorg{Person} or
\termsorg{Organization} class that performed the action is specified via the \termsorg{agent} property.
Fig~\ref{fig:process_crate_er} shows the classes used in Process Run Crate together with their relationships.

As an example,
suppose a user named John Doe runs the UNIX command \texttt{head} to extract the first ten lines of an input file named \texttt{lines.txt}, storing the result in another file called \texttt{selection.txt}.
John then runs the \texttt{sort}
UNIX command on \texttt{selection.txt}, storing the sorted output in a new file named \texttt{sorted\_selection.txt}.

\begin{figure}[!ht]
\includegraphics[width=29em]{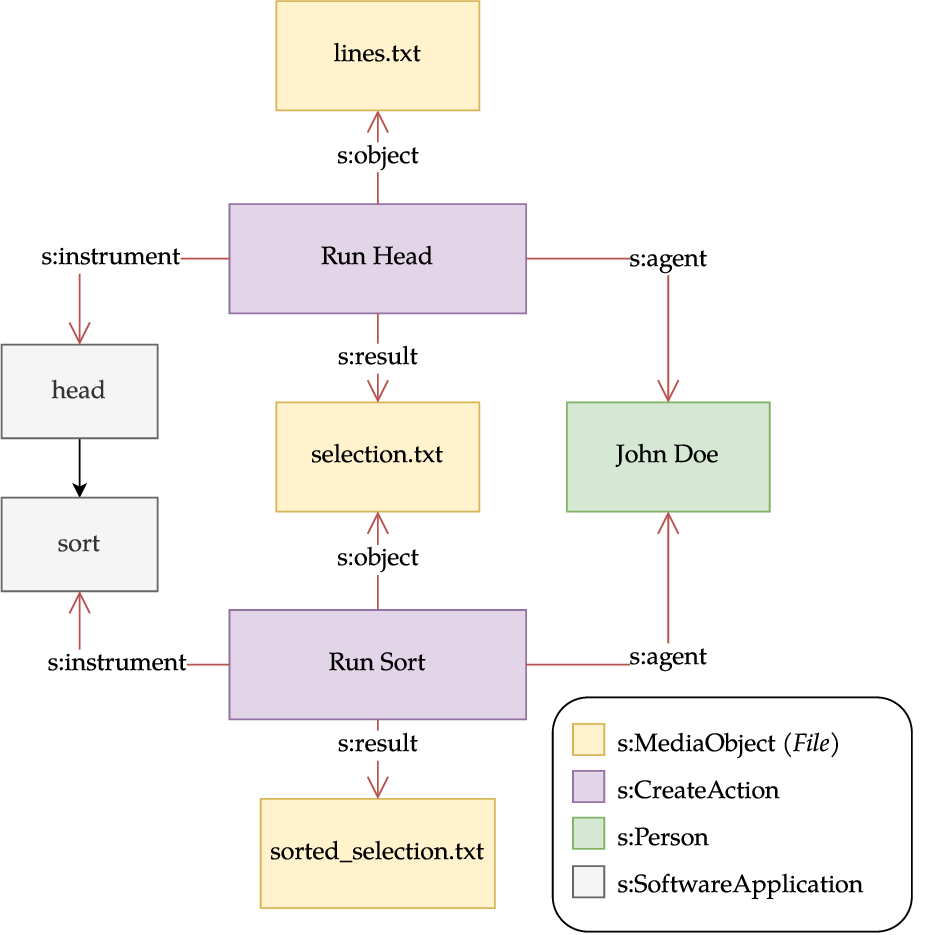}
\caption{{\bf Diagram of a simple workflow} where the \texttt{head} and \texttt{sort} programs were run manually by a user.
The executions of the individual software programs are connected by the fact that the file output by \texttt{head} was used as input for \texttt{sort}, documenting the computational flow in an implicit way.
Such executions can be represented with Process Run Crate.
}
\label{fig:head_sort}
\end{figure}

Fig~\ref{fig:head_sort} contains a diagram of the two actions and their relationships to the other involved entities.
Note how the actions are connected by the fact that the output of ``Run Head'' is also the input of ``Run Sort'': they form an ``implicit workflow'', whose steps have been executed manually rather than by a software tool.

Process Run Crate extends the RO-Crate guidelines on representing software used to create files with additional requirements and conventions.
This arrangement is typical of the RO-Crate approach, where the base specification provides general recommendations to allow for high flexibility, while profiles -- being more concerned with the representation of specific domains and machine actionability -- provide more detailed and structured definitions.
Nevertheless, in order to be broadly applicable, profiles also need to avoid the specification of too many strict requirements, trying to strike a good trade-off between flexibility and actionability.

\subsection{Workflow Run Crate}\label{workflow-run-crate}

\begin{figure}[!htb]
\includegraphics[width=26em]{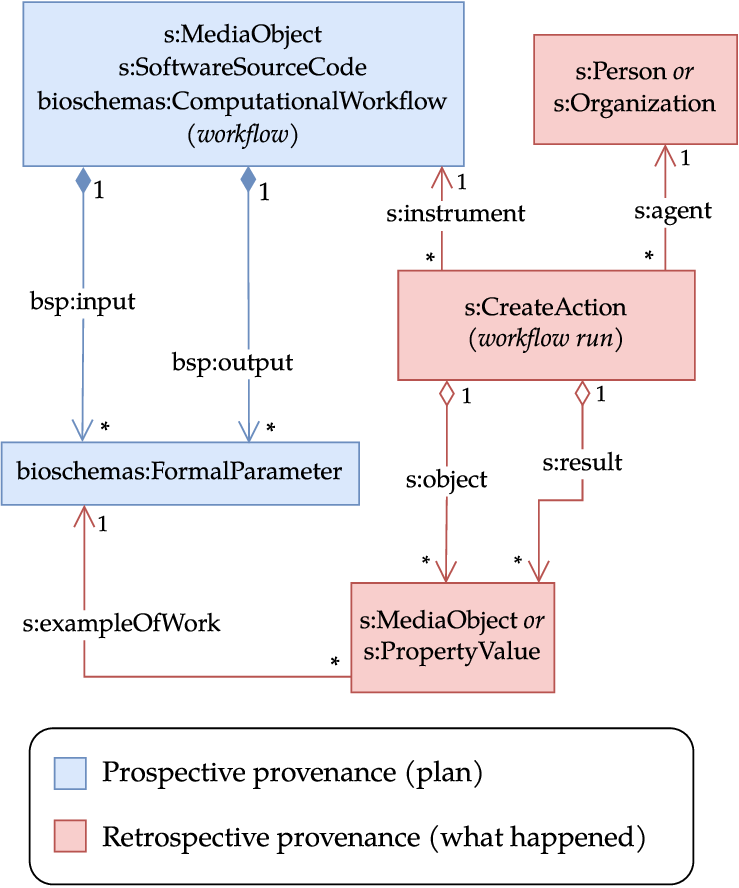}
\caption{{\bf UML class diagram for Workflow Run Crate.}
The main differences with Process Run Crate are the representation of formal parameters and the fact that the workflow is expected to be an entity with types \termsorg{MediaObject} (\emph{File} in RO-Crate JSON-LD), \termsorg{SoftwareSourceCode} and \termbioschemas{ComputationalWorkflow}.
Effectively, the workflow belongs to all three types, and its properties are the union of the properties of the individual types.
In this profile, the execution history (retrospective provenance) is augmented by a (prospective) workflow definition, giving a high-level overview of the workflow and its input and output parameter definitions (\termbioschemas{FormalParameter}). The inner structure of the workflow is not represented in this profile.
In the provenance part, individual files (\termsorg{MediaObject}) or arguments (\termsorg{PropertyValue}) are then connected to the parameters they realise. Most workflow systems can consume and produce multiple files, and this mechanism helps to declare each file's role in the workflow execution.
The filled diamond $\blacklozenge$ indicates composition, empty diamond $\Diamond$ aggregation, and other arrows relations.
}
\label{fig:workflow_crate_er}
\end{figure}

The Workflow Run Crate profile~\cite{WRROC 2024b} combines the Process Run Crate and WorkflowHub's Workflow RO-Crate~\cite{Bacall 2022} profiles to describe the execution of computational workflows managed by a WMS.
Such workflows are typically written in a domain-specific language, such as CWL or Snakemake
\cite{Koster 2012}, and run by one or more WMS (e.g., StreamFlow~\cite{Colonnelli 2021}, Galaxy~\cite{Galaxy 2022}).
Fig~\ref{fig:workflow_crate_er} illustrates the classes used in this profile together with their relationships.
As in Process Run Crate, the execution is described by a \termsorg{CreateAction}
that links to the application via \termsorg{instrument}, but in this case the application must be a workflow, as prescribed by Workflow RO-Crate.
More specifically, Workflow RO-Crate states that the RO-Crate must contain a main workflow typed as \emph{File} (an RO-Crate mapping to \termsorg{MediaObject}), \termsorg{SoftwareSourceCode}
and \termbioschemas{ComputationalWorkflow}.
The execution of the individual workflow steps, instead, is not represented: that is left to the more detailed Provenance Run Crate profile (described in the next section).

The Workflow Run Crate profile also contains recommendations on how to represent the workflow's input and output parameters, based on the Bioschemas ComputationalWorkflow profile.
All these elements are represented via the \termbioschemas{FormalParameter} class and are referenced from the main workflow via the \termbsp{input} and
\termbsp{output} properties.
While the classes referenced from
\termsorg{object} and \termsorg{result} in the \termsorg{CreateAction} represent data entities and argument values that were actually used in the workflow execution, the ones referenced from \termbsp{input} and
\termbsp{output} correspond to formal parameters, which acquire a value when the workflow is run (see Fig~\ref{fig:workflow_crate_er}).
In the profile, the relationship between an actual value and the corresponding formal parameter is expressed through the \termsorg{exampleOfWork} property.
For instance, in the following JSON-LD snippet a formal parameter (\texttt{\#annotations}) is illustrated together with a corresponding \texttt{final-annotations.tsv} file:
\begin{verbatim}
{
    "@id": "#annotations",
    "@type": "FormalParameter",
    "additionalType": "File",
    "encodingFormat": "text/tab-separated-values",
    "valueRequired": "True",
    "name": "annotations"
},
{
    "@id": "final-annotations.tsv",
    "@type": "File",
    "contentSize": "14784",
    "exampleOfWork": {"@id": "#annotations"}
}
\end{verbatim}


\subsection{Provenance Run Crate}\label{provenance-run-crate}

The Provenance Run Crate profile~\cite{WRROC 2024c} extends Workflow Run Crate by adding new concepts to describe the internal details of a workflow run, including individual tool executions, intermediate outputs and related parameters.
Individual tool executions are represented by additional \termsorg{CreateAction} instances that refer to the tool itself via \termsorg{instrument} -- analogously to its use in Process Run Crate.
The workflow is required to refer to the tools it orchestrates through the \termsorg{hasPart} property, as suggested in the Bioschemas ComputationalWorkflow profile, though in the latter it is only a recommendation.

To represent the logical steps defined by the workflow, this profile uses \termsorg{HowToStep} -- i.e., “A step in the instructions for how to achieve a result”~\cite{howtostep-def}.
Steps point to the corresponding tools via the \termsorg{workExample} property and are referenced from the workflow via the \termsorg{step} property; the execution of a step is represented by a \termsorg{ControlAction} pointing to the
\termsorg{HowToStep} via \termsorg{instrument} and to the \termsorg{CreateAction}
entities that represent the corresponding tool execution(s) via
\termsorg{object}.
Note that a step execution does not coincide with a tool execution: an example where this distinction is apparent is when a step maps to multiple executions of the same tool over a list of inputs (e.g. the ``scattering'' feature in CWL).

\begin{figure}[!htb]
\includegraphics[width=\textwidth]{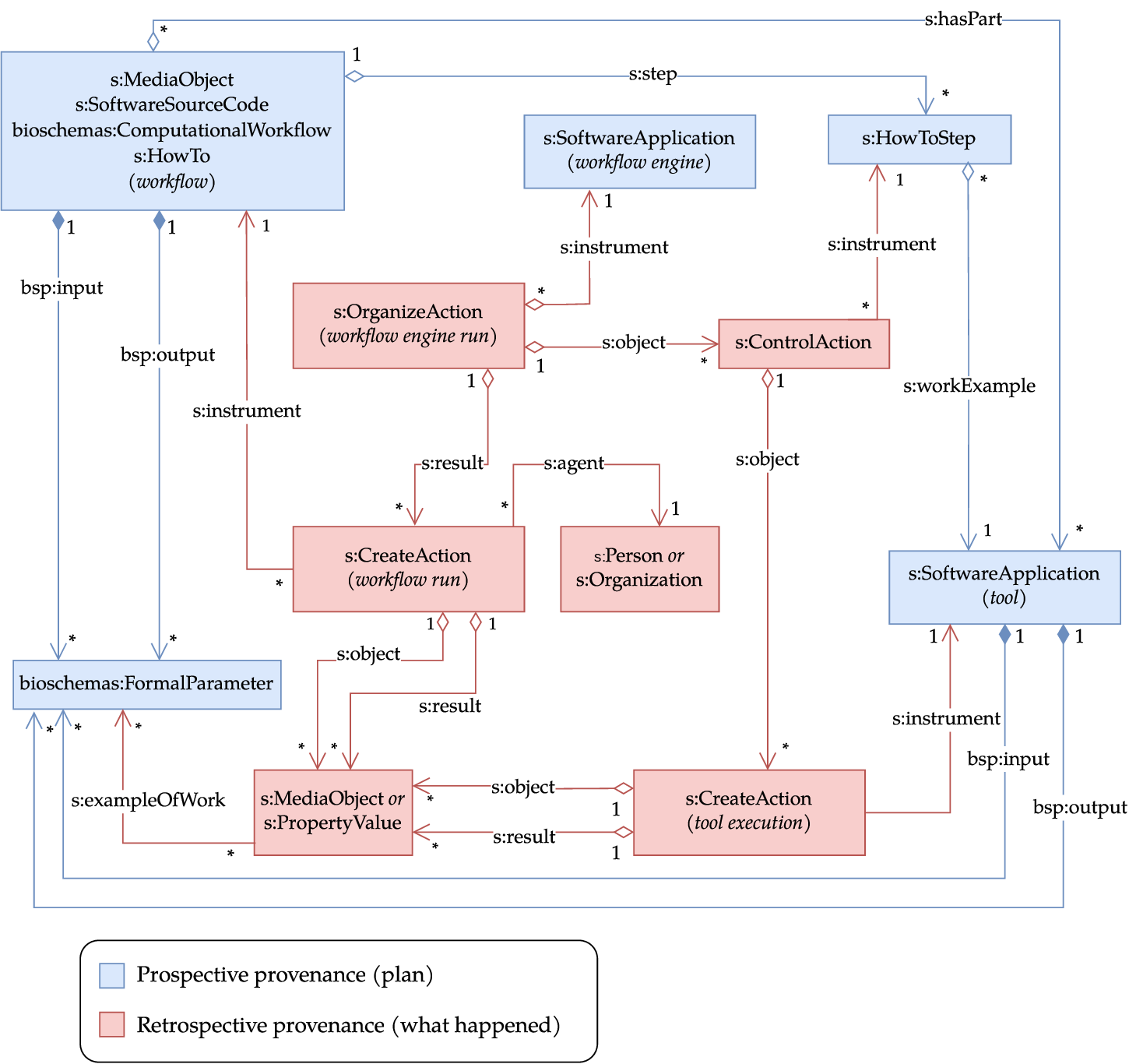}
\caption{{\bf UML class diagram for Provenance Run Crate.}
In addition to the workflow run, this profile represents the execution of individual steps and their related tools.
The prospective side (the execution plan) is shown by the workflow listing a series of \termsorg{HowToStep}s, each linking to the \termsorg{SoftwareApplication} that is to be executed. The \termbsp{input} and \termbsp{output} parameters for each tool are described in a similar way to the overall workflow parameter in Fig~\ref{fig:workflow_crate_er}.
The retrospective provenance side of this profile includes each tool execution as an additional \termsorg{CreateAction} with similar mapping to the realised parameters as \termsorg{MediaObject} or \termsorg{PropertyValue}, allowing intermediate values to be included in the RO-Crate even if they are not workflow outputs.
The workflow execution is described the same as in the Workflow Run Crate profile with an overall \termsorg{CreateAction} (the workflow outputs will typically also appear as outputs from inner tool executions). An additional \termsorg{OrganizeAction} represents the workflow engine execution, which orchestrated the steps from the workflow plan through corresponding \termsorg{ControlAction}s that spawned the tool's execution (\termsorg{CreateAction}). It is possible that a single workflow step had multiple such executions (e.g. array iterations). Not shown in figure: \termsorg{actionStatus} and \termsorg{error} to indicate step/workflow execution status.
The filled diamond $\blacklozenge$ indicates composition, empty diamond $\Diamond$ aggregation, and other arrows relations.
}
\label{fig:provenance_crate_er}
\end{figure}

An RO-Crate following this profile can also represent the execution of the WMS itself (e.g., cwltool) via
\termsorg{OrganizeAction}, pointing to a representation of the WMS via
\termsorg{instrument}, to the steps via \termsorg{object} and to the workflow run via \termsorg{result}.
The \termsorg{object} attribute of the
\termsorg{OrganizeAction} can additionally point to a configuration file containing a description of the settings that affected the behaviour of the WMS during the execution.
Fig~\ref{fig:provenance_crate_er} illustrates the various classes involved in the representation of a workflow run via Provenance Run Crate together with their relationships.

Additionally, this profile specifies how to describe connections between parameters,
through \textit{parameter connections} -- a fundamental feature of computational workflows.
Specifically, parameter connections describe: (i) how tools consume as input the intermediate outputs generated by other tools; and (ii) how workflow-level parameters are mapped to tool-level parameters.
As an example, consider again the workflow depicted in Fig~\ref{fig:head_sort},
and suppose it is implemented in a workflow language such as CWL: the workflow-level input (a text file) is linked through a parameter connection to the input of the \texttt{head} tool wrapper, and then a second parameter connection links this tool's output to the input of the \texttt{sort} tool wrapper.
A representation of parameter connections is particularly useful for traceability, since it provides the means to document the inputs and tools on which workflow outputs depend.
Since the current RO-Crate context has no suitable terms for the description of such relationships,
we added appropriate ones to the aforementioned dedicated term set~\cite{wrroc-terms}:
a \termwfrun{ParameterConnection} type with
\termwfrun{sourceParameter} and \termwfrun{targetParameter} attributes that respectively map to the source and target formal parameters, and a
\termwfrun{connection} property to link from the relevant step or workflow to the \termwfrun{ParameterConnection} instances.

\begin{figure}[htb]
  \includegraphics[width=26em]{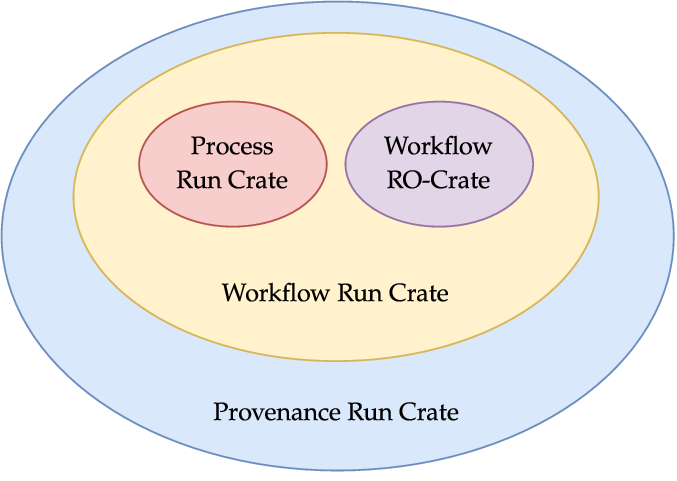}
  \caption{{\bf Venn diagram of the specifications for the various RO-Crate profiles.}
  Process Run Crate specifies how to describe the fundamental classes involved in a computational run, and thus is the basis for all profiles in the WRROC collection.
  Workflow Run Crate inherits the specifications of both Process Run Crate and Workflow RO-Crate. Provenance Run Crate, in turn, inherits the specifications of Workflow Run Crate (and in a sense includes multiple Process Runs for each step execution, but within a single Crate).
  }
  \label{fig:profile_venn}
  \end{figure}

In our set of profiles, Provenance Run Crate is the most detailed one and offers the highest level of granularity; its specification is a superset of Workflow Run RO-Crate, which in turn is a superset of Process Run Crate. This relationship between the three profiles is illustrated in Fig~\ref{fig:profile_venn}, as a Venn diagram.
Theoretically, all computational provenance information could be represented through the Provenance Run Crate profile alone (possibly relaxing some requirements), since it inherits from the other ones. In practice, though, this choice would require the use of the most complex model even for simple use cases. Having three separate profiles provides a way to represent information at different levels of granularity, while keeping all RO-Crates generated with them interoperable. This approach gives a straightforward path to supporting the representation of computational provenance in simpler use cases such as with simple command executions, i.e. the Process Run Crate. Additionally, the approach lowers the accessibility barrier for implementation in WMSs, as developers may choose to initially implement only the more basic support in their WMS, with reduced effort and complexity, and gradually scale to more detailed representations. This encourages the adoption of WRROC across the diverse landscape of use cases and WMSs.

\subsection{Profile formats}\label{profile-formats}

The WRROC profiles are available both in human-readable (HTML) and in machine-readable format (RO-Crate). The human-readable profiles are at:
\begin{itemize}
    \item \url{https://w3id.org/ro/wfrun/process/0.5}
    \item \url{https://w3id.org/ro/wfrun/workflow/0.5}
    \item \url{https://w3id.org/ro/wfrun/provenance/0.5}
\end{itemize}
And the corresponding machine-readable ones at:
\begin{itemize}
    \item \url{https://doi.org/10.5281/zenodo.12158562}
    \item \url{https://doi.org/10.5281/zenodo.12159311}
    \item \url{https://doi.org/10.5281/zenodo.12160782}
\end{itemize}
The RO-Crate metadata files for the machine readable profiles can be retrieved using the same URLs as the human-readable ones, but with JSON-LD content negotiation: this is done by setting \texttt{"Accept:application/ld+json"} in the HTTP header.

The new terms we defined to represent concepts that could not be expressed with existing Schema.org ones are at:
\begin{itemize}
    \item \url{https://w3id.org/ro/terms/workflow-run}
\end{itemize}
These terms are available in multiple formats with content negotiation, as explained at the above link.

\section{Implementations}\label{implementations}

Support for the Workflow Run RO-Crate profiles presented in this work has been implemented in a number of systems, showing support from the community and demonstrating their usability in practice.
We describe seven of these implementations (one in a conversion tool and six in WMS) in the following sections.
Table~\ref{implementation_summary_table} provides an overview of the implementations, along with the respective profile implemented, and links to the implementation itself and to an example RO-Crate.
These tools have been developed in parallel by different teams, and independently from each other.
RO-Crate has a strong ecosystem of tools~\cite{Soiland-Reyes 2022a}, and the WRROC implementations have either re-used these or added their own approach to the standards.

\subsection{Runcrate}\label{runcrate}

Runcrate~\cite{runcrate} is a Workflow Run RO-Crate toolkit which also serves as a reference implementation of the proposed profiles.
It consists of a Python package with a command line interface, providing a straightforward path to integration in Python software and other workflows.
The runcrate toolkit includes functionality to convert CWLProv ROs to RO-Crates conforming to the Provenance Run Crate profile (\texttt{runcrate convert}), effectively providing an indirect implementation of the format for cwltool.
Indeed, the CWLProv model provided a basis for the Provenance Run Crate profile, and the implementation of a conversion tool in runcrate at times drove the improvement and extension of the profile as new requirements or gaps in the old designs emerged.
Runcrate converts both the retrospective provenance part of the CWLProv RO (the RDF graph of the workflow's execution) and the prospective provenance part (the CWL files, including the workflow itself).
Both parts are thus converted into a single, workflow-language-agnostic metadata resource.

Another functionality offered by the runcrate package is \texttt{runcrate report}, which reports on the various executions described in an input RO-Crate, listing their starting and ending times, the values of the various parameters, etc.
Runcrate report demonstrates how the provenance profiles presented in this work enable comparison of runs interoperably across different workflow languages or different implementations of the same language.
This functionality has also been used as a lightweight validator for the various implementations.

Runcrate also includes a \texttt{run} subcommand to re-execute the computation described by an input Workflow Run Crate or Provenance Run Crate where CWL is used as a workflow language.
It works by mapping the RO-Crate description of input parameters and their values (the workflow's
\termbsp{input} and the action's \termsorg{object}) to the format expected by CWL, which is then used to relaunch the workflow on the input data.
This functionality shows the machine-actionability of the profiles to support reproducibility, and was used to successfully re-execute the digital pathology annotation workflow described in Section~\ref{provenance-run-crate-for-digital-pathology}.
Of course, achieving a full re-execution in the general case may not always be possible: reproducibility is supported by the profiles, but also benefits from specific characteristics of the workflow language (which should provide a clear formalism to map input items to their corresponding parameter slots) and of the specific workflow's implementation, which can be made considerably easier to reproduce by containerising the computational environment required by each step (if allowed by the workflow language).

\subsection{Galaxy}\label{galaxy}

The Galaxy project~\cite{Galaxy 2022} provides a WMS with data management functionalities as a multi-user platform, aiming to make computational biology more accessible to research scientists that do not have computer programming or systems administration experience.
Galaxy's most prominent features include: a collection of 7500+ integrated tools~\cite{Blankenberg 2014};
a web interface that allows the definition and execution of workflows using the integrated tools; a network of dedicated (public) Galaxy instances.

The export of workflow execution provenance data as Workflow Run Crates was added to Galaxy in version 23~\cite{Galaxy 2023} providing a more interoperable alternative to the basic export of Galaxy workflow
\emph{invocations}. A WRROC export from Galaxy includes: the workflow definition; a set of serialisations of the invocation-related metadata in Galaxy native, JSON-formatted files;
and the input and output data files.
This result is achieved by:
\begin{inlineenum}
\item extracting provenance data from Galaxy entities related to the workflow run, along with their associated metadata;
\item converting them to RO-Crate metadata using the ro-crate-py library~\cite{ro-crate-py};
\item describing all files contained in the basic invocation export within the RO-Crate metadata; and
\item making the Workflow Run Crate available for export to the user through Galaxy's web interface and API~\cite{De Geest 2022b}.
\end{inlineenum}
We extract the prospective provenance contained in Galaxy's YAML-based gxformat2
\cite{gxformat2} workflow definition, which includes details of the analysis pipeline such as the graph of the tools that need to be executed and metadata about the data types required.
The retrospective provenance -- i.e., the details of the executed workflow, such as the inputs, outputs, and parameter values used -- is extracted from Galaxy's data model,
which is not directly accessible to users in the context of a public Galaxy server.
All of this provenance information is then mapped to RO-Crate metadata, including some Galaxy-specific data entities such as dataset collections.
An exemplary Workflow Run Crate exported from Galaxy, through its \emph{Workflow Invocations} list, is available on Zenodo~\cite{De Geest 2023}.

In practice, a user would take the following steps to obtain a Workflow Run Crate from a Galaxy instance:
\begin{inlineenum}
\item create or download a Galaxy workflow definition (e.g.: from WorkflowHub) and import it in a Galaxy instance, or create a workflow through the Galaxy GUI directly;
\item execute the workflow, providing the required inputs;
\item after the workflow has run successfully, the corresponding RO-Crate will be available for export from the Workflow Invocations list.
\end{inlineenum}

\subsection{COMPSs}\label{compss}

COMPSs~\cite{Lordan 2014} is a task-based programming model that allows users to transform a sequential application into a parallel one by simply annotating some of its methods, thus facilitating scaling applications to increasing amounts of computing resources.
When a COMPSs application is executed, a corresponding workflow describing the application's tasks and their data dependencies is dynamically generated and used by the COMPSs runtime to orchestrate the execution of the application in the infrastructure.
As a WMS, COMPSs stands out for its many advanced features that enable applications to achieve fine-grained high efficiency in HPC systems, such as the ability to exploit underlying parallelisation frameworks (e.g.~MPI~\cite{Gabriel 2004}, OpenMP~\cite{Dagum 1998}), compilers (e.g.~NUMBA~\cite{Lam 2015}), failure management, task grouping, and more.  Also, provenance recording for COMPSs workflows has been explored in previous work~\cite{Sirvent 2022}, where the Workflow RO-Crate profile was used to capture structured descriptive metadata about the executed workflow, without introducing any significant run time performance overheads.

In this work, COMPSs has been further improved by implementing the generation of provenance
information conformant to the Workflow Run Crate profile, thus also capturing
details about the actual execution of the workflow.
The dynamic nature of COMPSs workflows poses some challenges to capturing
provenance, which were met thanks to the instruments
provided by the WRROC model.
For instance, a COMPSs workflow is created when the application is executed and, thus, a prior static workflow definition does not exist before that moment.
Due to this design, the workflow entity in the metadata file references the entry point application run by COMPSs -- instead of, for instance, a dedicated workflow definition file as one might find with other WMSs. Also, formal parameters are not included in the prospective provenance (note that specifying them is not required by the profile) because inputs and outputs (both for each task and the whole workflow) are determined at runtime.
However, the RO-Crate generation by COMPSs leverages the information recorded by the runtime to automatically add metadata of all input or output data assets used or produced by the workflow.

Because of the supercomputing environments where COMPSs is used, the integration of Workflow Run Crate support required paying particular attention to the generation of a unique ID for the \termsorg{CreateAction} representing the workflow run. Our implementation uses UUIDs for distributed environments, while it adds a combination of hostname and queuing system job ID for supercomputer executions, to provide as much information as possible from the run while preserving ID uniqueness.
In the \termsorg{CreateAction}, the \termsorg{description} term includes system information, as well as relevant environment variables that provide details on the execution environment (e.g., node list, CPUs per node).
Finally, the \termsorg{subjectOf} property of the \termsorg{CreateAction} references the system’s monitoring tool (when available),
where authorised users can see detailed profiling of the corresponding job execution, as provided by the MareNostrum IV supercomputer~\cite{marenostrum4-docs}.

To showcase the COMPSs adoption of the Workflow Run Crate profile, we provide as an example the execution of the BackTrackBB~\cite{Poiata 2016}
application in the MareNostrum IV supercomputer.
BackTrackBB targets the detection and location of seismic sources using the statistical coherence of the wave field recorded by seismic networks and antennas.
The resulting RO-Crate~\cite{Poiata 2023} captures the provenance of the execution results and complies with the Workflow Run Crate profile. It includes the application source files, a diagram of the workflow's graph, application profiling and input and output files.

The implementation of provenance recording using Workflow Run Crate has been fully integrated in the COMPSs runtime and is available as of release 3.2~\cite{Ejarque 2023}.

\subsection{StreamFlow}\label{streamflow}

The StreamFlow framework~\cite{Colonnelli 2021} is a container-native WMS for the execution of workflows defined in CWL.
It has been designed around two primary principles: first, it allows the execution of tasks in multi-container environments, supporting the concurrent execution of communicating tasks in a multi-agent ecosystem; second, it relaxes the requirement of a single shared data space, allowing for hybrid workflow executions on top of multi-cloud, hybrid cloud/HPC, and federated infrastructures.
StreamFlow orchestrates hybrid workflows by combining a \emph{workflow description} (e.g., a CWL workflow description and a set of input values) with one or more \emph{deployment descriptions} -- i.e.
representations of the execution environments in terms of infrastructure-as-code (e.g., Docker Compose files~\cite{Reis 2022}, HPC batch scripts, and Helm charts~\cite{Zerouali 2023}).
A \texttt{streamflow.yml} file -- the entry point of each StreamFlow execution -- binds each workflow step with the set of most suitable execution environments.
At execution time, StreamFlow automatically takes care of all the secondary aspects, like scheduling, checkpointing, fault tolerance, and data movements.

StreamFlow collects prospective and retrospective provenance data in a custom format and persists it into a pluggable database (using sqlite3 as the default choice).
After a CWL workflow execution completes, users can generate an RO-Crate through the \texttt{streamflow prov}
command, which extracts the provenance data stored in the database for one or more workflow executions and converts it to an RO-Crate archive that is fully compliant with the Provenance Run Crate Profile, including the details of each task run by the WMS.
Support for the format has been integrated into the main development branch and will be included in release 0.2.0~\cite{Colonnelli 2023b}.

From the StreamFlow point of view, the main limitation in the actual version of the Provenance Run Crate standard is the lack of support for distributed provenance -- i.e., a standard way to describe the set of locations involved in a workflow execution and their topology. As a temporary solution,
the StreamFlow configuration and a description of the hybrid execution environment are preserved by directly including the \texttt{streamflow.yml} file into the generated archive.
However, this product-specific solution prevents a wider adoption from other WMS and forces agnostic frameworks (e.g., WorkflowHub) to provide ad-hoc plugins to interpret the StreamFlow format.
Since the support for hybrid and cross-facility workflows is gaining traction in the WMS ecosystem, we envision support for distributed provenance as a feature for future versions of Workflow Run RO-Crate.

\subsection{WfExS-backend}\label{wfexs}

WfExS-backend~\cite{Fernandez 2024a} is a FAIR workflow execution orchestrator that aims to address some of the difficulties found in analysis reproducibility and analysis of sensitive data in a secure manner.
WfExS-backend requires that the software used by workflow steps is available in publicly accessible software containers for reproducibility.
Actual workflow execution is delegated to one of the supported workflow engines -- currently either Nextflow~\cite{Di Tommaso 2017} or cwltool.
The orchestrator prepares and stages all the elements needed to run the workflow -- i.e. all the files of the workflow itself, the specific version of the workflow engine, the required software containers and the inputs.
All these elements are referenced through resolvable identifiers, ideally public, permanent ones.
Thanks to this approach, the orchestrator can consume workflows from various types of sources, such as git repositories, Software Heritage, or even RO-Crates from WorkflowHub.
WfExS-backend development milestones have aimed to reach FAIR workflow execution through the generation and consumption of RO-Crates following the Workflow Run Crate profile, which has proven to be a mechanism suitable to semantically describe digital objects in a way that simplifies embedding details crucial to analysis reproducibility and replicability.

When the orchestrator prepares a workflow for execution it records details relevant to the prospective provenance, such as the public URLs used to fetch input data and workflows, content digestion fingerprints (typically sha256 checksums) and metadata derived from workflow files, container images and input files.
Most of this captured metadata is later included in the generated RO-Crates. WfExS-backend has explicit commands to generate and publish both prospective and retrospective provenance RO-Crates based on a given existing staged execution scenario.
These RO-Crates can selectively include copies of used elements as payloads.
Workflows can be executed more than once in the same staged directory, with all the executions sharing the same inputs.
In this case, run details from all the executions are represented in the retrospective provenance RO-Crate. Support for the consumption of Workflow Run RO-Crates to reproduce the operations they document is available as of WfExS-backend version 1.0.0a0~\cite{Fernandez 2024a}.
We have created examples of Workflow Run Crates generated by WfExS-backend to capture provenance information from the execution of a Nextflow workflow~\cite{Bouyssie 2023} and a CWL workflow~\cite{Amstutz 2023}; these crates are both available on Zenodo~\cite{Fernandez 2024b, Fernandez 2024c}.
Future developments to WfExS-backend will also add support for embedding in the RO-Crates the URLs of output results that have been deposited into a suitable repository (like Zenodo DOIs, for instance).

\subsection{Sapporo}\label{sapporo}

Sapporo~\cite{Suetake 2022a} is an implementation of the Workflow Execution Service (WES) API specification~\cite{Rehm 2021}.
WES is a standard proposed by the Global Alliance for Genomics and Health (GA4GH) for cloud-based data analysis platforms that receive requests to execute workflows.
Sapporo supports the execution of several workflow engines, including cwltool~\cite{Amstutz 2023}, Toil~\cite{Vivian 2017}, and StreamFlow~\cite{Colonnelli 2021}.
Sapporo includes features specifically tailored to bioinformatics applications, including the calculation of feature statistics from specific types of outputs generated by workflow runs.
For example, the system calculates the mapping rate of DNA sequence alignments from BAM format files.
To describe the feature values, Sapporo uses the Workflow Run Crate profile extended with additional terms to represent these biological features~\cite{sapporo-terms}.

Further, the Tonkaz companion command line software has integrated functionality to compare Run Crates generated by Sapporo to measure the reproducibility of the workflow outputs~\cite{Suetake 2023}.
Developers can use this unique feature to build a CI/CD platform for their workflows to ensure that changes to the product do not produce an unexpected result.
Workflow users can also use this feature to verify the results from the same workflow deployed in different environments.

While Sapporo supports Workflow Run Crate, since WES is a WMS wrapper, it does not parse the provided workflow definition files. 
Instead, it embeds the information in the files passed by the WES request to record the provenance of execution rather than using the actual workflow parameters meant for the wrapped WMS.
Therefore, the current implementation of Sapporo does not capture the connections between the inputs/outputs depicted in the workflow and the actual files used/generated during the run.
The profile generated by Sapporo has fields representing input and output files, but they are not linked to formal parameters.

Sapporo supports export to Workflow Run Crate as of release 1.5.1~\cite{Suetake 2023b}. An example of a Workflow Run RO-Crate generated by Sapporo is available on Zenodo~\cite{Ohta 2023}.

\subsection{Autosubmit}\label{autosubmit}

Autosubmit~\cite{Manubens-Gil 2016} is an open source, lightweight workflow manager and meta-scheduler tailored to configuring and running scientific experiments in climate research. It supports scheduling jobs via SSH to Slurm~\cite{Yoo 2003}, PBS~\cite{Feng 2007} and other remote batch servers used in HPC.

Autosubmit's ``archive'' feature archives the experiment directory and all its contents into a ZIP file, which can be used later to access the provenance data or to execute the Autosubmit experiment again.
Even though the data in the ZIP file includes prospective provenance and retrospective provenance, it is not structured, and a simple examination yields no way to distinguish the provenance types.

Recent releases of Autosubmit 4 have added features to increase user flexibility.  An updated YAML configuration management system has been implemented that allows users to combine multiple YAML files into a single unified configuration file.
Also, the option to use only the experiment manager features of Autosubmit has been added, delegating the workflow execution to a different backend workflow engine -- like ecFlow~\cite{Bahra 2011}, Cylc~\cite{Oliver 2019}, or a CWL runner.
While these features provide some much appreciated flexibility, they have increased the complexity involved in reliably tracking the experiment configuration and other metadata for provenance documentation purposes.

In order to give users a more structured way to archive provenance, which includes the complete experiment configuration, the parameters used to generate it, and is also interoperable between workflow managers, the archive feature was enhanced with a new option in Autosubmit 4.0.100~\cite{Beltran 2023} to enable the generation of provenance data in Workflow Run RO-Crates.
The prospective provenance data for the crate is extracted from the Autosubmit experiment configuration.
This data includes the multiple YAML files, the unified YAML configuration, as well as the parameters used to preprocess each file -- preprocessing replaces placeholders in script templates with values from the experiment configuration.
The retrospective provenance data is included with the RO-Crate archive and includes logs and other traces produced by the experiment workflow.
Both prospective and retrospective provenance data are included in the final RO-Crate, which is compliant with the Workflow Run Crate profile.
At a practical level, the implementation was able to leverage the \texttt{ro-crate-py} library for many of the details pertaining to the creation of the RO-Crate archive in Python, and adding information for the JSON-LD metadata.

One of the main challenges for implementing WRROC support in Autosubmit was incorporating Autosubmit's \emph{Project} feature.
A Project in Autosubmit is an abstract concept that references a code repository and is used to define experiment configuration and contains template scripts defining workflow tasks and other auxiliary files.
%
%
The project has a type that defines the \emph{type} of the repository (e.g., git) and a \emph{location} that is the URL to retrieve it.
The RO-Crate file generated by Autosubmit includes the project type and location, but it does not include the complete Project and so it is lacking configuration details and scripts.
Therefore, users receive provenance data of the Project, but only those with the appropriate privileges can access its constituent resources (many applications run with Autosubmit can not be publicly shared without consent).
After consulting with the RO-Crate community regarding the specific Autosubmit requirements, the Autosubmit team adopted a mixed approach where Autosubmit initialises the JSON-LD metadata from its configuration and local trace files, and the user is responsible for providing a partial JSON-LD metadata object in the Autosubmit YAML configuration.
\texttt{ro-crate-py} was extended to allow the RO-Crate JSON-LD metadata to be patched by these partial JSON-LD metadata objects.
This way, users are able to provide the information that is missing from the Autosubmit configuration model, but is required by WRROC -- e.g., licence, authors, inputs, outputs, formal parameters, etc.

Future implementations of WRROC support should be facilitated by the new
functionality added to ro-crate-py to support the user-mediated metadata
integration approach.
%
On the other hand, the integration of WRROC support would have been facilitated by an automated validation tool for RO-Crate archives, and by documentation and examples on how to use the profiles with \emph{coarse-grained} workflow management systems (as defined in~\cite{Goble 2020}) that do not track inputs and outputs, which is the case of Autosubmit -- as well as the Cylc and ecFlow workflow engines.  The feedback generated by this use case was welcomed by the WRROC community and work to address these issues is either planned on under way at the time of writing.

To demonstrate Autosubmit's new WRROC-based functionality to generate structured provenance data, a workflow was created using an example Autosubmit Project designed using UFZ's mHM (mesoscale Hydrological Model)~\cite{Samaniego 2010,Kumar 2013}, and it was executed with Autosubmit. The resulting Workflow Run Crate is available from Zenodo~\cite{Kinoshita 2023}.

\begin{table}[htb]
  \begin{adjustwidth}{-1.4cm}{0in} 
  \centering
  \caption{
  {\bf Workflow Run Crate implementations}}
  \begin{tabular}{l|l|l|l}
  \hline
  {\bf Impl.} & {\bf Profile} & {\bf Version URL/DOI} &
  {\bf Example}\\
  \thickhline
  runcrate & Provenance & \footnotesize \cite{runcrate}  & \footnotesize \cite{run-pathology} \\
  Galaxy & Workflow & \footnotesize \cite{Galaxy 2023} & \footnotesize \cite{De Geest 2023} \\
  COMPSs & Workflow & \footnotesize \cite{Ejarque 2023} & \footnotesize \cite{Poiata 2023} \\
  Streamflow & Provenance & \footnotesize \cite{Colonnelli 2023b} & \footnotesize \cite{Colonnelli 2023} \\
  WfExS & Workflow & \footnotesize \cite{Fernandez 2024a} & \footnotesize \cite{Fernandez 2024b} \\
  Sapporo & Workflow & \footnotesize \cite{Suetake 2023b} & \footnotesize \cite{Ohta 2023} \\
  Autosubmit & Workflow & \footnotesize \cite{Beltran 2023} & \footnotesize \cite{Kinoshita 2023} \\
  \end{tabular}
  \begin{flushleft} 
    Summary of each WRROC implementation, together with the profile it implements, the software version that makes it available and an example RO-Crate. Runcrate is a toolkit that converts CWLProv ROs to Provenance Run Crates, while the others are workflow management systems.
  \end{flushleft}
  \label{implementation_summary_table}
  \end{adjustwidth}
\end{table}

\section{Exemplary use cases}\label{exemplary-use-cases}

We illustrate Workflow Run RO-Crate on two exemplary use cases. These are similar in terms of application domain, as they both relate to the application of machine learning techniques for the analysis of human prostate images for the purpose of supporting cancer tissue detection. However, the use cases are quite different in the way computations are executed and provenance is represented: in the first, the analysis is conducted by means of a CWL workflow and the outcome is represented with Provenance Run Crate; in the second, Process Run Crate is used in combination with a complementary model to represent a provenance chain that can extend beyond the computational analysis.

\subsection{Provenance Run Crate for digital pathology}\label{provenance-run-crate-for-digital-pathology}

In this section, we present a use case that demonstrates the effectiveness of the Provenance Run Crate profile at capturing provenance data in the context of digital pathology.
More specifically, we demonstrate the generation of RO-Crates to save provenance data associated with the computational annotation of magnified prostate tissue areas and cancer subregions using deep learning models~\cite{Del Rio 2022}.
The image annotation process is implemented in a CWL workflow consisting of three steps, each executing inference on an image using a deep learning model: \begin{inlineenum}
\item inference of a low-resolution tissue mask to select areas for further processing;
\item high-resolution tissue inference to refine borders;
\item high-resolution cancer tissue identification.
\end{inlineenum}
The two tissue inference steps run the same tool, but set different values for the parameter that controls the magnification level, and the second runs on a subset of the image area.
The workflow is integrated in the CRS4 Digital Pathology Platform~\cite{digital-pathology-platform}, a web-based platform to support clinical studies involving the examination and/or the annotation of digital pathology images.

To assess the interoperability of WRROC, we recorded the provenance of the execution of the same exemplary workflow on two different WMSs.
In the first case, we executed the CWL workflow with cwltool and converted the resulting CWLProv RO to a Provenance Run Crate with the runcrate tool (Section~\ref{runcrate}).
In the second case, the workflow was executed with the StreamFlow WMS (Section~\ref{streamflow}).
The RO-Crates obtained in the two cases~\cite{run-pathology, Colonnelli 2023}
are very similar to each other, differing only in a few details. For instance, Streamflow includes its configuration file in the crate and has separate files for the workflow and the two tools, while
cwltool with runcrate results in the workflow and the tools being stored in a single file (CWL's ``packed'' format).
Apart from these minor differences, the description of the computation is essentially the same, so the RO-Crates are fully interoperable.
Four actions are represented: the workflow itself, the two executions of the tissue extraction tool and the execution of the tumour classification tool.
Each action is linked to the corresponding workflow or tool via the
\termsorg{instrument} property, and reports its starting and ending time. For each action, input and output slots are referenced by the workflow, while the corresponding values are referenced by the action itself.
The data and \termsorg{PropertyValue} entities corresponding to the input and output values link to the corresponding parameter slots via the \termsorg{exampleOfWork} property, providing information on the values taken by the parameters during execution.
Listing~\ref{lst:ml_pipeline_streamflow_report} shows the output of the
\texttt{runcrate report} command for the StreamFlow RO-Crate. 
For each action (workflow or tool run), runcrate reports the associated instrument (workflow or tool), the starting and ending time and the list of inputs and outputs, with pointers from the formal parameter to the corresponding actual value taken during the execution of the action.

\begin{lstlisting}[float,basicstyle=\scriptsize\ttfamily,caption={Output of the \texttt{runcrate report} command executed on the Provenance Run Crate generated by StreamFlow in the digital pathology inference use case (Section~\ref{provenance-run-crate-for-digital-pathology}). This informal listing of relevant RO-Crate entities describes each step of the execution. Note that inputs and outputs are of different types (not shown): e.g., \texttt{tissue\_low>0.9} is a string parameter, \texttt{6b15de\dots} is a filename, and \texttt{\#af0253\dots} is a collection.},label={lst:ml_pipeline_streamflow_report}]
action: #30a65cba-1b75-47dc-ad47-1d33819cf156
  instrument: predictions.cwl (['SoftwareSourceCode', 
         'ComputationalWorkflow', 'HowTo', 'File'])
  started: 2023-05-09T05:10:53.937305+00:00
  ended: 2023-05-09T05:11:07.521396+00:00
  inputs:
    #af0253d688f3409a2c6d24bf6b35df7c4e271292 <- predictions.cwl#slide
    tissue_low <- predictions.cwl#tissue-low-label
    9 <- predictions.cwl#tissue-low-level
    tissue_low>0.9 <- predictions.cwl#tissue-high-filter
    tissue_high <- predictions.cwl#tissue-high-label
    4 <- predictions.cwl#tissue-high-level
    tissue_low>0.99 <- predictions.cwl#tumor-filter
    tumor <- predictions.cwl#tumor-label
    1 <- predictions.cwl#tumor-level
  outputs:
    06133ec5f8973ec3cc5281e5df56421c3228c221 <- predictions.cwl#tissue
    4fd6110ee3c544182027f82ffe84b5ae7db5fb81 <- predictions.cwl#tumor
action: #457c80d0-75e8-46d6-bada-b3fe82ea0ef1
  step: predictions.cwl#extract-tissue-low
  instrument: extract_tissue.cwl (['SoftwareApplication', 'File'])
  started: 2023-05-09T05:10:55.236742+00:00
  ended: 2023-05-09T05:10:55.910025+00:00
  inputs:
    tissue_low <- extract_tissue.cwl#label
    9 <- extract_tissue.cwl#level
    #af0253d688f3409a2c6d24bf6b35df7c4e271292 <- extract_tissue.cwl#src
  outputs:
    6b15de40dd0ee3234062d0f261c77575a60de0f2 <- extract_tissue.cwl#tissue
action: #d09a8355-1a14-4ea4-b00b-122e010e5cc9
  step: predictions.cwl#extract-tissue-high
  instrument: extract_tissue.cwl (['SoftwareApplication', 'File'])
  started: 2023-05-09T05:10:58.417760+00:00
  ended: 2023-05-09T05:11:03.153912+00:00
  inputs:
    tissue_low>0.9 <- extract_tissue.cwl#filter
    6b15de40dd0ee3234062d0f261c77575a60de0f2 <- extract_tissue.cwl#filter_slide
    tissue_high <- extract_tissue.cwl#label
    4 <- extract_tissue.cwl#level
    #af0253d688f3409a2c6d24bf6b35df7c4e271292 <- extract_tissue.cwl#src
  outputs:
    06133ec5f8973ec3cc5281e5df56421c3228c221 <- extract_tissue.cwl#tissue
action: #ae2163a8-1a2a-4d78-9c81-caad76a72e47
  step: predictions.cwl#classify-tumor
  instrument: classify_tumor.cwl (['SoftwareApplication', 'File'])
  started: 2023-05-09T05:10:58.420654+00:00
  ended: 2023-05-09T05:11:06.708344+00:00
  inputs:
    tissue_low>0.99 <- classify_tumor.cwl#filter
    6b15de40dd0ee3234062d0f261c77575a60de0f2 <- classify_tumor.cwl#filter_slide
    tumor <- classify_tumor.cwl#label
    1 <- classify_tumor.cwl#level
    #af0253d688f3409a2c6d24bf6b35df7c4e271292 <- classify_tumor.cwl#src
  outputs:
    4fd6110ee3c544182027f82ffe84b5ae7db5fb81 <- classify_tumor.cwl#tumor
\end{lstlisting}

The \termsorg{exampleOfWork} link between input / output values and parameter slots is used by \texttt{runcrate run} to reconstruct the CWL input parameter mapping needed to rerun the computation.
The \termsorg{alternateName} property (a Schema.org property applicable to all entities), which records the original name of data entities (at the time the computation was run), is also crucial for reproducibility in this case: both StreamFlow and CWLProv, to avoid clashes, record input and output files and directories using their SHA1
checksum as their names. 
However, for this particular workflow file names are important: it expects the input image data to be in the MIRAX~\cite{mirax-format} format, where the ``main'' dataset file taken as an input parameter by the processing application must be accompanied by a directory of additional data files, in the same location and with the same name, apart from the extension.
The runcrate tool uses the \termsorg{alternateName} to rename the input dataset as required, so that the expected pattern can be picked up by the workflow during the re-execution.
This use case was the main motivation to include a recommendation to use \termsorg{alternateName} with the above semantics in Process Run Crate.

Thanks to the fact that both RO-Crates were generated following the best practices to support reproducibility mentioned in the profiles, we were able to automatically re-execute both computations with the runcrate tool.
This was also made possible by the fact that the CWL workflow included information on which container images to use for each tool.
Overall, this shows how reproducibility is a hard-to-achieve goal that can only be supported, but not ensured, by the profiles, since it also depends on factors like the characteristics of the computation, the choice of workflow language and whether best practices such as containerisation are followed.

This use case highlighted the need to add specifications on how to represent multi-file datasets~\cite[section ``Representing multi-file objects"]{WRROC 2024a}, driven by the need to handle the aforementioned MIRAX image format.
To represent these, we added specifications to the Process Run Crate profile on describing “composite” datasets consisting of multiple files and directories to be treated as a single unit -- as opposed to more conventional input or output parameters consisting of a single file. The profile specifies that such datasets should be represented by a \termsorg{Collection} class linking to individual files and directories via the \termsorg{hasPart} property, and referencing the main part (if any) via the \termsorg{mainEntity} property. Note that, by adding this specification to Process Run Crate, we also made it available to Workflow Run Crate and Provenance Run Crate. In the output of the runcrate report tool the additional files are not shown, since the formal parameter points to the \termsorg{Collection} class that describes the whole dataset.

This use case also demonstrates the usage of parameter connections (described in Section~\ref{provenance-run-crate}). The RO-Crate resulting from the workflow run contains a representation of all connections between workflow-level parameters (the overall input and output parameters) and tool-level parameters. This allows crate consumers to programmatically find which tool is affected by a workflow-level parameter, thus providing insight on how the workflow works internally (the main feature of the Provenance Run Crate profile). For instance, the \texttt{tissue-high-level} workflow parameter is connected to the \texttt{level} parameter of the \texttt{extract\_tissue.cwl} tool by the \texttt{extract-tissue-high} step. This parameter regulates the resolution level (pyramidal images are organised into multiple levels of resolution) at which the image is processed in the high-resolution tissue extraction phase. A similar connection is present for the tissue extraction at low resolution. Since \termwfrun{ParameterConnection}s are referenced from the relevant \termsorg{HowToStep}, the crate consumer can easily determine the resolution level used for both image processing phases from the retrospective provenance.

\subsection{Process Run Crate and CPM RO-Crate for cancer detection}\label{process-run-crate-and-cpm-ro-crate-for-cancer-detection}

This section presents an RO-Crate created to describe an execution of a computational pipeline that trains AI models to detect the presence of carcinoma cells in high-resolution digital images of magnified human prostate tissue.
This RO-Crate makes use of Process Run Crate and CPM RO-Crate~\cite{cpm-ro-crate}, an RO-Crate profile that supports the representation of entities described according to the Common Provenance Model (CPM)~\cite{Wittner 2022,Wittner 2023b, Wittner 2024}.

The CPM is a recently developed extension of the W3C PROV model~\cite{Moreau 2013}. It enables the representation of distributed provenance,
which is created when an object involved in the research process -- either digital or physical (e.g., biological material) -- is exchanged between organisations, so that each organisation can document only a portion of the object’s life cycle.
Using CPM, each involed organisation can document its portion of the life cycle by generating, storing, and managing individual provenance components, which are then linked together in a chain that spans multiple organizations.
The CPM prescribes how to represent such provenance, and how to enable its traversal and processing using a common algorithm, independently from the type of object being described. In addition, the CPM defines a notion of meta-provenance, which contains metadata about the history of individual provenance components.

CPM RO-Crate supports the identification of CPM-based provenance and meta-provenance files within an RO-Crate, so that data, metadata, and CPM-based provenance information can be packed together.
An RO-Crate generated according to the CPM-RO-Crate profile embeds parts of the distributed provenance, which may be linked to the provenance of precursors and successors of the packed data.
The CPM-RO-Crate profile synergises well with Process Run Crate, since the former can add references to CPM-based provenance descriptions of computational executions described with the latter, integrating them in the distributed provenance. Since CPM-based provenance and meta-provenance files are typically themselves produced by computations, Process Run Crate allows to represent these along with the main computations that produce the datasets being exchanged, providing the full picture in a cohesive ensemble.

The use case pipeline consists of three main computational steps:
\begin{inlineenum}
\item a preprocessing step that splits input images into small patches and divides them into a training and a testing set;
\item a training step that trains the model to recognise the presence of carcinoma cells in the images;
\item an evaluation step that measures the accuracy of the trained model on the testing set.
\end{inlineenum}
In addition to these pipeline steps, the RO-Crate describes additional computations related to the generation of the CPM provenance and meta-provenance files.
All computations are described according to the Process Run Crate profile, while the CPM files are referenced according to the CPM RO-Crate profile. 
Also represented via Process Run Crate are: the input dataset; the results of the pipeline execution; the scripts that implement the pipeline; the log files generated by the scripts; a script that converts the logs into the CPM files.
This approach allowed us to describe all elements as a single RO-Crate, which
is available on Zenodo~\cite{Wittner 2023a}.

Listing~\ref{lst:model_training_pipeline_report} presents the \texttt{runcrate report} output for the RO-Crate,
including action inputs and outputs while omitting other details. The listing shows the connections between the actions, forming an ``implicit workflow'' as discussed in Section~\ref{process-run-crate}. For instance, the \texttt{prov\_train.log} file is both an output of the training action (\texttt{\#train\_script:ROCRATE-PUB-\ldots}) and an input of the CPM provenance generation action for the training phase (\texttt{\#train\_script:6efa9a06-\ldots:CPM-provgen}), highlighting the interdependency between the steps.

\begin{lstlisting}[float,basicstyle=\scriptsize\ttfamily,caption={Excerpt of the output of the \texttt{runcrate report} command for the AI model training Process Run Crate; only inputs and outputs of the actions are shown. The listing shows the connections between the pipeline actions through the entities they produce or consume -- e.g., \texttt{cam16\_mrxs.h5} is output of the conversion script \texttt{convert\_script:ff67\ldots} and input for the training script \texttt{train\_script:ROCRATE\ldots}},label={lst:model_training_pipeline_report}]
action: #convert_script:ff67ce65-736f-46d5-9fec-10953cad8695
  inputs:
    wsi/test/
    wsi/train/
    prov_converter_config.json
  outputs:
    cam16_mrxs.h5
    prov_preprocess.log

action: #test_script:ROCRATE-PUB-1438b57a750ce887d4433d9e
  inputs:
    prov_test_config.json
    cam16_mrxs.h5
  outputs:
    predictions.h5
    prov_test.log

action: #test_script:d3cfd9cf-6851-43c6-bee9-c8dc18f22368:CPM-provgen
  inputs:
    prov_test.log
  outputs:
    prov_test.provn
    prov_test.provn.log
    prov_test.png

action: #train_script:ROCRATE-PUB-1438b57a750ce887d4433d9e
  inputs:
    prov_train_config.json
    cam16_mrxs.h5
  outputs:
    prov_train.log
    model/weights/auc_01.ckpt.index
    model/weights/auc_01.ckpt.data-00000-of-00001
    model/weights/auc_02.ckpt.index
    model/weights/auc_02.ckpt.data-00000-of-00001
    model/weights/best_loss.ckpt.index
    model/weights/best_loss.ckpt.data-00000-of-00001
    model/weights/auc_03.ckpt.index
    model/weights/auc_03.ckpt.data-00000-of-00001

action: #train_script:6efa9a06-b8e9-4cfc-88c7-e9d35e5263c3:CPM-provgen
  inputs:
    prov_train.log
  outputs:
    prov_train.provn
    prov_train.png
    prov_train.provn.log

action: #convert_script:9d030b68-70d8-4526-82fe-160d9cfe4806:CPM-provgen
  inputs:
    prov_preprocess.log
  outputs:
    prov_preprocess.provn
    prov_preprocess.png
    prov_preprocess.provn.log

action: #meta_provn_script:86bae258-4c51-4215-854b-32cb49f239ab:CPM-provgen
  inputs:
    prov_train.provn.log
    prov_test.provn.log
    prov_preprocess.provn.log
  outputs:
    meta_provenance.provn
    meta_provenance.png
    meta_provenance.provn.log
\end{lstlisting}

The CPM files complement the RO-Crate with details about the pipeline execution process, such as how the input dataset was split into training and testing sets, or detailed information about each training iteration of the AI model.
For instance, the RO-Crate contains a representation of a checkpoint of the AI model after the second training iteration, with the corresponding entity's attributes containing paths to the respective model stored as a file.
The entity is related to the respective training iteration activity, which contains the iteration parameters represented as an attribute list.
In addition, the CPM generally provides means to link the input dataset provenance to the provenance of its precursors -- human prostate tissues and biological samples the tissues were derived from; this is not included in the example because we used a publicly available input database for which provenance of the precursors was not available.
However, the linking mechanism for provenance precursors is exactly the same as between the bundles for the AI pipeline parts.
While the RO-Crate is focused on the execution of the pipeline, the provenance included in the CPM files intends to be interlinked with provenance of the precursors or successors, providing means to traverse the whole provenance chain.
For the described digital pathology pipeline, the precursors would be:
\begin{inlineenum}
\item a biological sample acquired from a patient;
\item slices of the sample processed and put on glass slides;
\item the images created as a result of scanning the slides using a microscope.
\end{inlineenum}
As a result, combining the CPM and RO-Crate enables the lookup of research artefacts related to the computation across heterogeneous organisations using the underlying provenance chain.

\section{Discussion}\label{discussion}

The RO-Crate profiles presented in this work provide a unified data model to describe the prospective and retrospective provenance of the execution of a computational workflow, together with contextual metadata about the workflow itself and its associated entities (inputs, outputs, code, etc.).
The profiles are flexible, allowing one to tailor the provenance description to a broad variety of use cases, agnostic to the WMS used, and allow describing provenance traces at different levels of granularity.
These characteristics facilitate implementing support in workflow systems. Six WMS have already integrated support for a WRROC profile, as described in Section~\ref{implementations}. These new RO-Crate profiles enable interoperability between implementations, which has been demonstrated through the comparison of workflow executions on heterogeneous systems.

Choosing to base our approach on the RO-Crate model has led to a number of
benefits. The collected provenance data can be treated with standard RDF tools. As an example, the following SPARQL~\cite{sparql11-overview} query returns all actions in a Workflow Run RO-Crate, together with their instruments and their starting and ending times, independently of the original workflow type or the WMS that executed the workflow:
\begin{verbatim}
PREFIX schema: <https://schema.org/>
SELECT ?action ?instrument ?start ?end
WHERE {
  ?action a schema:CreateAction .
  ?action schema:instrument ?instrument .
  OPTIONAL { ?action schema:startTime ?start } .
  OPTIONAL { ?action schema:endTime ?end }
}
\end{verbatim}
Further, having workflow runs and plans described according to the RO-Crate model allows capturing the context of the workflow itself (e.g.~authors, related publications, other workflows, etc.), in addition to the trace alone.
%
Another advantage of RO-Crate is that the files corresponding to the data entities (inputs, outputs, code, etc.) do not necessarily have to be stored together with the metadata file: for instance, they can be remote and referred to via an http(s) URI. This aspect is mostly relevant in situations where the file is very large or cannot be shared publicly, since a URI can reference a resource to which access is limited (e.g., accessible only after authentication, or from specific network boundaries, etc.).

The WRROC profiles are extensions of the base RO-Crate specification that specialise it for the use case of workflow execution provenance representation. The additional terms, constraints and recommendations introduced by the profiles allow users to represent classes and relationships involved in a workflow execution in a precise and detailed way, so that consumers of the RO-Crate can programmatically retrieve the relevant information according to predefined patterns and act upon it. This is a crucial advantage over using the base RO-Crate specification, which was not designed to answer the competency questions defined for capturing the provenance of workflow executions.

The ability to build FAIR into Workflow Management Systems was identified as one of the current open challenges in the Scientific Workflows domain at
the Workflows Community Summit~\cite{Ferreira 2023}, with the objective of achieving FAIR Computational Workflows. The profiles introduced in this article help tackle this challenge by introducing interoperable metadata among WMSs that captures the provenance of their corresponding workflow executions.
The derivation of Workflow Run Crate, and in turn Provenance Run Crate, from Workflow RO-Crate makes the digital objects that conform to these new profiles compatible with the WorkflowHub workflow registry~\cite{Goble 2021}. This design entails that Workflow Run RO-Crates directly reference the workflow with which the provenance was generated, and it allows workflow runs to be registered on WorkflowHub and easily found and shared with other researchers. Additionally, the inheritance mechanism allows reusing the specifications already developed for Workflow RO-Crate, which form part of the guidelines on representing the prospective provenance.

The Workflow Run RO-Crate profiles, the associated tooling, the implementations and the examples are developed and supported by the open WRROC Community.
At the time of writing, the Community numbers nearly 40 members and brings together members of the RO-Crate community~\cite{Soiland-Reyes 2022a}, WMS users and developers, workflow users and developers, GA4GH~\cite{Rehm 2021} Cloud developers and provenance model authors, and is open to anyone who is interested in the representation of workflow execution provenance.
The inclusion of WMS developers and workflow users has been key to keeping the specifications flexible, easy to implement and grounded on real use cases, while the diversity of the stakeholders has included a plurality of viewpoints while driving the model's development forward, resulting in profiles that are already being used (as described in Section~\ref{implementations}).

In the following subsections, we provide an evaluation of the metadata coverage of runcrate and we discuss how WRROC relates to standards such as W3C PROV-O and to other community projects.

\subsection{Evaluation of metadata coverage using runcrate convert}

Since CWLProv was a starting point in the development of WRROC (Section~\ref{runcrate}), as a baseline validation we chose to verify that the metadata contained in CWLProv ROs is preserved in the RO-Crates produced by their conversion through runcrate's \emph{convert} command. In previous work we had conducted a qualitative analysis of metadata coverage in CWLProv (version 0.6.0), based on concrete examples of ROs associated with a realistic bioinformatics workflow~\cite{De Wit 2022};
in this work we repeated this analysis for WRROC, and compared the WRROC RDF representation (in \texttt{ro-crate-metadata.json}) with the CWLProv RDF provenance graph.
To summarise, the analysis focuses on the comparison of the degree of representation by the two models of six provenance data
types defined in~\cite{De Wit 2022}, which we recall here for clarity.
\begin{enumerate}[label={\bfseries T\arabic*.}]
  \item {\bf Scientific context}: the choices which were made in the design of the workflow and parameter values.
  \item {\bf Data}: input and output data.
  \item {\bf Software}: the tools directly orchestrated by the workflow, and their dependencies.
  \item {\bf Workflow}: the workflow and tool descriptions, but not the software they control.
  \item {\bf Computational environment}: metadata about the system on which the workflow was executed, comprising both software and hardware.
  \item {\bf Execution details}: additional information about the workflow execution itself.
\end{enumerate}
Each type is in turn articulated in a set of data subtypes, forming a hierarchy
of elements that should be represented in
workflow provenance data to satisfy a range of use cases spanning from
supporting workflow development to supporting a service based on the
execution of the workflow, with several other use cases in between.  For a full
motivation and description of the criteria the reader may refer to the original work~\cite{De Wit 2022}.

Our analysis shows that, overall, most of the information contained in the CWLProv RDF is transferred to the RO-Crate metadata.
The results are summarised in Table~\ref{analysis_table};
for completeness, we also report the (non-RDF) representation of provenance metadata in CWL-specific documents (\texttt{packed.cwl} and \texttt{primary-job.json}), which are included in both CWLProv ROs and RO-Crates generated by runcrate.
We observe that out of the total 20 provenance data subtypes that are part of the analysis, WRROC represented 13 (65\%) of them (9 fully, 4 partially), while CWLProv RDF captured 8 (3 fully, 5 partially).  The representation of some entire categories of metadata has improved -- notably Workflow parameters (WF2), which were insufficiently described in CWLProv RDF, but defined with type and format in RO-Crate.
Moreover, the Workflow Run RO-Crate RDF contains a representation of tools orchestrated by the workflow (T3), as well as a much more extensive description of the workflow itself (T4) compared to CWLProv.

In conclusion, our analysis shows that runcrate preserves most provenance metadata previously shown to be relevant in realistic RO use case scenarios.
More detailed results of the analysis can be found in~\cite{de Wit 2024}. 

\begin{table}[ht]
\begin{adjustwidth}{-5cm}{0in} 
\centering
\caption{
{\bf Summarised results of our qualitative analysis of Provenance Run Crates generated with runcrate.}}
\begin{tabular}{c|r|l|l|c|c}
\hline
{\bf CWL (non-RDF)} & {\bf Type} & {\bf Subtype}      & {\bf Name} & {\bf CWLProv RDF} & {\bf WRROC RDF}  
\\ \thickhline
$\bullet$ & T1 & SC1 & Workflow design  &   $\cdot$ & $\bullet$  \\ 
$\circ$ & & SC2 & Entity annotations      &  $\cdot$ &  $\cdot$   \\ 
$\cdot$  & & SC3 & Workflow execution ann. &  $\cdot$ &  $\cdot$  \\ \hline
$\circ$ & T2 & D1 & Data identification   & $\cdot$ &  $\cdot$ \\
$\circ$ & & D2 & File characteristics     & $\circ$ & $\circ$ \\
$\circ$ & & D3 & Data access              &  $\cdot$ &  $\cdot$  \\ 
$\bullet$ & & D4 & Parameter mapping        & $\bullet$ & $\bullet$ \\ \hline 
$\bullet$ & T3 & SW1 & Software identification &  $\cdot$ & $\bullet$  \\ 
$\bullet$ & & SW2 & Software documentation  &  $\cdot$ & $\bullet$  \\  
$\bullet$ & & SW3 & Software access         &  $\cdot$ & $\bullet$ \\ \hline 
$\bullet$ & T4 & WF1 & Workflow software    & $\circ$ & $\bullet$  \\ 
$\bullet$ & & WF2 & Workflow parameters     & $\circ$ & $\bullet$  \\ 
$\bullet$ & & WF3 & Workflow requirements   &  $\cdot$  &  $\circ$  \\ \hline 
$\cdot$ & T5 & ENV1 & Software environment & $\cdot$ &  $\cdot$  \\ 
$\cdot$ & & ENV2 & Hardware environment   & $\cdot$ &  $\cdot$  \\ 
$\circ$ & & ENV3 & Container image        & $\circ$ &  $\circ$  \\ \hline 
$\cdot$ & T6 & EX1 & Execution timestamps & $\bullet$ & $\bullet$ \\ 
$\cdot$ & & EX2 & Consumed resources      &  $\cdot$ & $\cdot$  \\ 
$\cdot$ & & EX3 & Workflow engine         & $\circ$ & $\circ$  \\  
$\cdot$ & & EX4 & Human agent             & $\bullet$ & $\bullet$  \\ \hline
\end{tabular}
\begin{flushleft}
  We converted CWLProv (v0.6.0) ROs to WRROC with runcrate 0.5.0. The table compares the
  degree to which the data subtypes of the provenance data taxonomy
  (identified by the triple (\texttt{Type}, \texttt{Subtype}, \texttt{Name})) are preserved
  by the CWLProv RDF and the WRROC RDF models; the taxonomy is defined in previous work~\cite{De Wit 2022},
  where relevant provenance metadata are identified based on realistic
  use cases for ROs associated with a real-life bioinformatics workflow.
For completeness, the \emph{CWL (non-RDF)} column also reports the non-RDF representation of provenance metadata
in CWL-specific documents: \texttt{packed.cwl} (the workflow) and \texttt{primary-job.json} (the input parameter file).
Since \texttt{packed.cwl} and \texttt{primary-job.json} are also included in RO-Crate, we only considered how the metadata was represented in \texttt{ro-crate-metadata.json}. \\
\textbf{Legend:} $\bullet$ fully represented  $\;\;\circ$ partially represented   $\;\;\cdot$ missing or unstructured representation 
\end{flushleft}
\label{analysis_table}
\end{adjustwidth}
\end{table}

\subsection{Workflow Run RO-Crate and the W3C PROV standard}

One of our aims for the WRROC profiles is to make them compatible with both Schema.org and W3C PROV. Provenance Run Crate is the profile that most closely matches the level of detail provided by CWLProv, which extends W3C PROV. Table~\ref{rocrate_prov_mapping} shows how the main classes and relationships represented by Provenance Run Crate map to PROV constructs, using the SKOS vocabulary to indicate the type of relationship between each pair of terms. A machine-readable version of the mapping can be found in the RO-Crate accompanying this article~\cite{wrroc-crate,wrroc-crate-html}.

\begin{table}[h]
  \begin{adjustwidth}{-1.5cm}{0in}
  \centering
  \caption{
  {\bf Mapping from Workflow Run RO-Crate to equivalent W3C PROV concepts} using SKOS~\cite{Isaac 2009}. For instance, \termsorg{CreateAction} has \textbf{broader} match \emph{prov:Activity}, meaning that \emph{prov:Activity} is more general. Prefix \emph{prov:} \url{https://www.w3.org/ns/prov\#}.}
  \begin{tabular}{p{60mm}|p{40mm}|p{40mm}}
  \hline
  {\bf RO-Crate} & \textbf{Relationship} & {\bf W3C PROV-O} \\
  \thickhline

  \termsorg{Action} (superclass of \termsorg{CreateAction}, \termsorg{OrganizeAction}) &
    Has close match
    \begin{small}
      (Schema.org Actions may also be potential actions in the future)
    \end{small}
    &
    \emph{prov:Activity}
    \\ \hline
  \termsorg{CreateAction}, \termsorg{OrganizeAction} &
    Has broader match &
    \emph{prov:Activity}
    \\ \hline
  \termsorg{Person} &
    Has exact match &
    \emph{prov:Person}
    \\ \hline
  \termsorg{Organization} &
    Has exact match &
    \emph{prov:Organization}
    \\ \hline
  \termsorg{SoftwareApplication} &
    Has related match &
    \emph{prov:SoftwareAgent}
    \\ \hline
  \termbioschemas{ComputationalWorkflow}, \termsorg{SoftwareApplication}, \termsorg{HowTo} &
    Has broader match &
    \emph{prov:Plan},
    \emph{prov:Entity}
    \\ \hline
  \termsorg{MediaObject}, \termsorg{Dataset}, \termsorg{PropertyValue} &
    Has broader match &
    \emph{prov:Entity}
    \\ \hline
  \termsorg{startTime} on \termsorg{CreateAction} &
    Has close match &
    \emph{prov:startedAtTime}
    \\ \hline
  \termsorg{endTime} on \termsorg{CreateAction} &
    Has close match &
    \emph{prov:endedAtTime}
    \\ \hline
  \termsorg{agent} on \termsorg{CreateAction} &
    Has related match &
    \emph{prov:wasStartedBy}, \emph{prov:wasEndedBy}
    \\ \hline
  \termsorg{agent} and \termsorg{instrument} on \termsorg{CreateAction} &
    Has broader match &
    \emph{prov:wasAssociatedWith}
    \\ \hline
  \termsorg{instrument} on \termsorg{CreateAction} &
    Has related match
    \begin{small}
      (Complex mapping: an instrument implies a qualified association with the agent, linked to a plan)
    \end{small}
    &
    \emph{prov:hadPlan} on \emph{prov:Association}
    \\ \hline

  \termsorg{object} on \termsorg{CreateAction} &
    Has exact match &
    \emph{prov:used}
    \\ \hline
  \termsorg{result} on \termsorg{CreateAction} &
    Has close match &
    inverse \emph{prov:wasGeneratedBy}

  \end{tabular}
  \label{rocrate_prov_mapping}
  \end{adjustwidth}
\end{table}

\subsection{Five Safes Workflow Run Crate}\label{trusted-workflow-run-crate}

The \emph{Five Safes RO-Crate}~\cite{5s-crate} profile has been developed to extend the Workflow Run Crate profile for use in Trusted Research Environments (TRE), following the Five Safes Framework~\cite{Desai 2016} to better handle sensitive health data in federated workflow execution across TREs in the UK~\cite{trefx}.
A crate with a workflow run request references a pre-approved workflow and project details for manual and automated assessment according to the TRE's agreement policy for the sensitive dataset.
The crate then goes through multiple phases internal to the TRE, including validation, sign-off, workflow execution and disclosure control.
At this stage the crate is also conforming to the Workflow Run Crate profile.
The final crate is then safe to be made public.

This extension of Workflow Run Crate documents and supports the \emph{human review process} -- important for transparency on TRE data usage. 
The initial implementation of this process used WfExS as the workflow execution backend, and this approach will form the basis for further work on implementing federated workflow execution in the British initiatives DARE UK and HDR UK~\cite{Snowley 2023} and in the European EOSC-ENTRUST project for Trusted Research Environments~\cite{eosc-entrust}.

\subsection{Biocompute Object RO-Crate}\label{bco-crate}
IEEE 2791-2020~\cite{Mazumder 2020}, colloquially known as \emph{Biocompute Objects} (BCO), is a standard for representing provenance of a genomic sequencing pipeline, intended for submission of the workflow to regulatory bodies -- e.g. as part of a personalised medical treatment method~\cite{Alterovitz 2018}.
The BCO is represented as a single JSON file which includes description of the workflow and its steps and intended purpose, as well as references for tools used and data sources accessed. 
There is overlap in the goals of BCO and Workflow Run Crate profiles, however their intentions and focus are different. 
BCO is primarily conveying a computational method for the purpose of manual regulatory review and further reuse, with any values provided as an exemplar run.  
A Workflow Run Crate, however, is primarily documenting a particular workflow execution, and the workflow is associated to facilitate rerun rather than reuse. 

Previously, a guide to packaging BioCompute Objects using RO-Crate~\cite{bco-roc} was developed as a profile to combine both standards~\cite{Soiland-Reyes 2021}.
In this early approach, RO-Crate was primarily a vessel to transport the BCO along with its constituent resources, including the workflow and data files, as well as to provide these resources with additional typing and licence metadata that is not captured by the BCO JSON.
Further work is being planned with the BCO community to update the BCO-RO profile to align with the newer Workflow Run RO-Crate profiles. 

\hypertarget{conclusion}{%
\section{Conclusion and future work}\label{conclusion}}

The Workflow Run RO-Crate profile collection presented in this manuscript is a
new model to represent and package both the prospective
and the retrospective provenance relating to the execution of computational
workflows in a way that is
machine-actionable, interoperable, independent of the specific workflow language or
execution system, and including support for re-execution.
These new profiles build on RO-Crate and Schema.org to include contextual
information and bundle together all objects of the workflow execution
(inputs, outputs, code, etc.).
Our approach minimizes the set of mandatory metadata items
and defines a hierarchy of profiles -- Process Run Crate, Workflow Run
Crate, and Provenance Run Crate -- that capture provenance information at increasing
levels of detail and complexity.
This flexible approach increases the model’s adaptability to the diverse
landscape of WMSs used in practice, and modulates the implementation effort as a
function of the requirements of the specific use case.
As a result, there has already been significant uptake of Workflow Run RO-Crate, as shown by its adoption in six WMS, including Galaxy, StreamFlow and COMPSs;
in addition, the \texttt{runcrate} toolkit has been implemented as part of this
work providing various inspection, conversion and re-execution functionalities.
Moreover, we have shown how WRROC has been applied in real use cases.

Workflow Run RO-Crate is an ongoing project. Therefore, our profiles and the surrounding software are not static entities, but keep being updated to cater for new requirements and use cases.
As examples of ongoing work, at the time of writing there are plans to expand the runcrate toolkit to better support the creation and querying of WRROC objects.  Also, work is ongoing to implement automated conformance validation of crates.
In addition, several of the implementations presented in this work will also develop new features. For instance, the Galaxy community plans to extend its WRROC support to: include metadata detailing each step of a workflow run to conform to the Provenance Run Crate profile; develop and/or integrate RO-Crate more deeply with import and export of Galaxy histories; and further develop user-guided import of RO-Crates as Galaxy datasets, histories and workflows.
Further, we are currently exploring the cloud execution of Workflow Run RO-Crates.
The Workflow Execution Service (WES) specification is used by the Global Alliance for Genomics and Health (GA4GH)~\cite{Rehm 2021} to enable WMS-agnostic interpretation of workflows and scheduling of task execution. In addition, the Task Execution Service (TES) specification enables the execution of individual, atomic, containerised tasks in a compute backend-independent manner.
We are planning to undertake an in-depth analysis of the degree of interoperability between the TES and WES API standards -- roughly the equivalents of Process and Workflow Run Crates, respectively -- by placing their focus on the actual execution of tasks/processes and workflows in cloud environments and liaising with the GA4GH Cloud community to align schemas where necessary.
We will then build an interconversion library that attempts to
\begin{inlineenum}
\item construct WES workflow and TES task run requests from RO-Crates containing Provenance, Workflow or Process Run requests and therefore allow their easy (re)execution on any GA4GH Cloud API-powered infrastructure, and
\item bundle information from the WES and TES (as well as other GA4GH Cloud API resources, where available) to create or extend RO-Crates with standards-compliant Process, Workflow or even Provenance RO-Crates.
\end{inlineenum}

The maintenance and development of WRROC is driven by an open community,
currently numbering about 40 members. The Community runs regular virtual
meetings (every two weeks at the time of writing) and coordinates on Slack and
the RO-Crate mailing list.
Naturally, feedback and contributions from the community are welcome and
encouraged, and new requirements and features are discussed and sustained, particularly
through the WRROC GitHub repository issue tracker~\cite{run-crate-repository}.
Through the open Community we expect to encourage and support further adoption of WRROC, be it by the other WMS or other use cases, maybe in time converging towards a common workflow execution provenance representation.



\section*{Acknowledgments}

The authors would like to thank all participants to the Workflow Run
RO-Crate working group meetings for the fruitful discussions and
valuable feedback.

\subsection*{Author contributions}
Author contributions following the CRediT Taxonomy:

\begin{description}
\item[Simone Leo]
Conceptualization, Data Curation, Investigation, Methodology, Resources, Software, Supervision, Validation, Visualization, Writing -- Original Draft preparation, Writing -- Review \& Editing
\item[Michael R. Crusoe]
Conceptualization, Investigation, Software, Supervision
\item[Laura Rodríguez-Navas]
Software, Writing -- Original Draft preparation
\item[Raül Sirvent]
Data Curation, Software, Writing -- Original Draft preparation, Writing -- Review \& Editing
\item[Alexander Kanitz]
Writing -- Original Draft preparation, Writing -- Review \& Editing
\item[Paul De Geest]
Data Curation, Software, Writing -- Original Draft preparation
\item[Rudolf Wittner]
Data Curation, Writing -- Original Draft preparation, Writing -- Review \& Editing
\item[Luca Pireddu]
Funding acquisition, Project Administration, Supervision, Writing -- Review \& Editing
\item[Daniel Garijo]
Conceptualization, Formal Analysis, Writing -- Original Draft preparation, Writing -- Review \& Editing
\item[José M. Fernández]
Data Curation, Software, Writing -- Original Draft preparation
\item[Iacopo Colonnelli]
Data Curation, Software, Writing -- Original Draft preparation
\item[Matej Gallo]
Data Curation, Software
\item[Tazro Ohta]
Data Curation, Software, Writing -- Original Draft preparation
\item[Hirotaka Suetake]
Data Curation, Software, Writing -- Original Draft preparation
\item[Salvador Capella-Gutierrez]
Funding Acquisition, Resources, Supervision, Writing -- Original Draft preparation
\item[Renske de Wit]
Software, Writing -- Original Draft preparation, Writing -- Review \& Editing
\item[Bruno de Paula Kinoshita]
Data Curation, Software, Writing -- Original Draft preparation, Writing -- Review \& Editing
\item[Stian Soiland-Reyes]
Conceptualization, Formal Analysis, Funding Acquisition, Investigation, Methodology, Resources, Software, Supervision, Visualization, Writing -- Original Draft preparation, Writing -- Review \& Editing
\end{description}

\nolinenumbers

%
%
%

\end{document}